\documentclass{article}
\usepackage{amssymb,amsmath}
\newcommand{\p}{{\partial}}
\newcommand{\PSbox}[3]{\mbox{\rule{0in}{#3}\hspace{#2}\includegraphics{#1}}}

\newtheorem{thm}{Theorem}
\newtheorem{cor}[thm]{Corollary}
\newtheorem{lemma}[thm]{Lemma}
\newtheorem{prop}[thm]{Proposition}

\newenvironment{proof}{{\bf Proof. }}{\hfill$\square$ \vskip .5cm}

\newcommand{\R}{{\mathbb R}}
\newcommand{\C}{{\mathbb C}}
\newcommand{\Z}{{\mathbb Z}}
\newcommand{\N}{{\mathbb N}}

\renewcommand{\Re}{\mbox{Re}}
\renewcommand{\Im}{\mbox{Im}}
\newcommand{\eps}{{\epsilon}}

\begin{document}
\title{The asymptotic determinant of the discrete Laplacian}
\author{Richard Kenyon
\thanks{CNRS UMR8628, Laboratoire de Topologie et Dynamique, B\^at. 425,
Universit\'e Paris-Sud, 91405 Orsay, France. email kenyon@topo.math.u-psud.fr}}
\maketitle

\abstract{
We compute the asymptotic determinant of 
the discrete Laplacian on a simply-connected
rectilinear region in $\R^2$. Specifically, for each $\eps>0$
let $H_\eps$ be the subgraph of $\eps\Z^2$ whose vertices lie in a fixed
rectilinear polygon $U$. Let $N(H_\eps)$ denote the number of vertices of $H_\eps$
and $B(H_\eps)$ the number of vertices on the boundary (the outer face).
Then the log of the determinant of the Laplacian on $H_\eps$
has the following asymptotic expansion in $\eps$:
$$\frac{4G}\pi N(H_\eps)+\frac{\log(\sqrt2-1)}2B(H_\eps) 
-\frac\pi{48}r_2(\eps,U) + o(1)$$
where $G$ is Catalan's constant and 
$r_2(\eps,U)$, which is $O(\log\frac1\eps)$, is the Dirichlet 
energy of a certain canonical harmonic function $h$ on $U$.

As an application of this result, we
prove that the growth exponent of the loop-erased random walk in $\Z^2$
is $5/4$.
}

\begin{center}
{\bf R\'esum\'e}\end{center}
{\small Nous calculons le d\'eveloppement asymptotique du 
d\'eterminant du Laplacien 
discret sur une r\'egion rectilin\'eaire de $\R^2$. 
Comme application, nous montrons que l'\'esp\`erance de la longueur
de la marche al\'eatoire laplacienne dans $\Z^2$ 
est \'a croissance $n^{5/4}$.}

\section{Introduction}
The determinant of the Laplacian on a graph arises in two related
statistical mechanical models, the uniform spanning tree model
and the two-dimensional lattice dimer model. For the spanning tree model,
Kirchhoff \cite{Kirch} is attributed with showing that 
the product of the non-zero eigenvalues of the Laplacian on a finite
graph is the same as the number of spanning trees on that graph.
For the two-dimensional dimer model, Temperley \cite{Temp}, 
based on work of Kasteleyn \cite{Kast}, 
showed that the number of dimer coverings of certain subgraphs of $\Z^2$ 
can be computed by the determinant of the Laplacian on related graphs.
Precise estimates on these determinants provide important
information about these models, in particular allowing one to compute
certain critical exponents and correlation functions \cite{Dup,Kenyon.prob}.

In this paper we compute the asymptotic expansion of the determinant
of the Laplacian on a special family of graphs: subgraphs of $\Z^2$
which are approximating rectilinear polygons (a polygon
is {\bf rectilinear} if its sides are parallel to the axes).
Our main motivation is not to study the Laplacian in itself
but rather to study both the dimer (domino tiling) 
model and the uniform spanning
tree model. For this reason we use the language of domino tilings.
(Domino tilings are tilings with $1\times 2$ and $2\times 1$ rectangles.)

Temperley \cite{Temp} gave a bijection between the number of spanning
trees of a subgraph $H$ of $\Z^2$ and domino tilings of a polyomino $P=P(H)$
constructed from the superposition of $H$ and its dual. We call a polyomino 
{\bf Temperleyan} if it arises from a graph $H$ by Temperley's construction.
Such polyominos have a simple description:
let $Q$ be a polyomino such that each side between a
concave and convex corner has even length, and each side between
two concave or two convex corners has odd length. Such a polyomino
has odd area; let $P$ be obtained from $Q$ by removing one lattice 
square at some convex corner. Then $P$ is Temperleyan. 

\begin{thm} \label{main2} Let $U\subset\R^2$ be a rectilinear polygon
with $V$ vertices.  For each
$\eps>0$, let $P_\eps$ be a Temperleyan polyomino in $\eps\Z^2$ approximating
$U$ in the natural sense (the corners of $P_\eps$ are converging to the corners
of $U$). Let $A_\eps$ be the area and $Perim_\eps$ be the perimeter of $P_\eps$.
Then the log of the number of domino tilings of $P_\eps$ is
\begin{equation}\label{main2form}
\frac{c_0A_\eps}{\eps^2} + \frac{c_1\mbox{Perim}_\eps}\eps- \frac\pi{48}\left(c_2(\eps)\log\frac1\eps+c_3(U)\right)+c_4 + o(1),
\end{equation}
where $c_0=\frac{G}\pi$, $G=1-\frac1{3^2}+\frac1{5^2}-\dots$ is Catalan's constant, 
$c_1=\frac{G}{2\pi}+\frac{\log(\sqrt{2}-1)}4$, $c_4$ is a constant independent of $U$
and
$c_2(\eps)\log\frac1\eps + c_3(U)$ is the $\eps$-normalized Dirichlet
energy of the limiting average height function on $U$ (see definitions below).
The term $-\frac\pi{48}c_2(\eps)$ is of the form
$$-\frac12-\frac{V-4}{36} \left(1+{ERR}(\eps)\right)$$
where ${ERR}(\eps)$ is $o(1)$.
\end{thm} 

\begin{cor}\label{main1} Under the above hypotheses on $U$,
for each $\eps>0$ let $H_\eps$ be the subgraph of $\eps\Z^2$ 
whose vertices are in $U$. Let $N(H_\eps)$ be the number of vertices in $H_\eps$
and $B(H_\eps)$ the number of edges of $\eps\Z^2$ on the boundary of $H_\eps$.
Then the log of the determinant of the Laplacian on $H_\eps$ is 
$$\frac{4G}\pi N(H_\eps)+ \frac{\log(\sqrt2-1)}2B(H_\eps)-\frac\pi{48}\left(c_2(\eps)\log\frac1\eps
+c_3(U)\right)+ c_5 + o(1),$$
where $c_2,c_3$ are as in Theorem \ref{main2} and $c_5$ is another constant independent of $U$.
\end{cor}

The limiting average height function has the following description.
Let $b_0\in \partial U$ be a base point.
For $x\in\partial U$ define $u_0(x)$ to be the total turning (in radians)
of the boundary tangent on the boundary path counterclockwise from $b_0$
to $x$ (the function 
$u_0$ has jump discontinuities at each corner of $U$ and at $b_0$).
The limiting average height function is the harmonic function on $U$ whose
boundary values are $\frac2\pi u_0$.
The $\eps$-normalized Dirichlet energy is by definition 
the Dirichlet energy contained
in the complement of the $\eps$-neighborhoods of the jump discontinuities.

\noindent{\bf Remarks}

\noindent{\bf 1.} In the case $U$ is a rectangle, formula (\ref{main2form})
follows from the formula
for the exact number of tilings computed by Kasteleyn \cite{Kast} and Temperley and
Fisher \cite{TF} (the asymptotic expansion of which was 
computed by Duplantier and David \cite{Dup}); see Proposition \ref{epsrectform} below.
\smallskip

\noindent{\bf 2.}
The leading term in the above formula, involving the constant $c_0$,
essentially follows from work of Burton and Pemantle \cite{BP}:
They constructed
a measure $\mu$ of entropy $c_0$ on the space $X$ of domino tilings of the plane
and proved that it was the unique translation-invariant 
measure of maximal entropy on $X$. Furthermore they proved
that for regions of the type used in the theorem, the entropy (the
coefficient of $\eps^{-2}$ in (\ref{main2form})) is $c_0A$.
\smallskip

\noindent{\bf 3.}
Our boundary conditions give rise to a correction to the number
of tilings which is only exponential in the length of the boundary
(the `perimeter' term in the theorem).
In \cite{CKP}, on the contrary,
it was shown that in some sense ``most'' other boundary 
conditions have a larger effect, giving a smaller entropy $c_0$, and
making the local densities of configurations vary throughout the region.
So both the `Area' and `Perimeter' terms in the theorem
depend strongly on our choice of boundary conditions.
\smallskip

\noindent{\bf 4.}
Note that if two regions have the same area, perimeter and number of vertices
then the log of the number of domino tilings differs by a constant in the limit,
that is, the ratio of the number of tilings is tending to a constant as $\eps\to0$.
This constant depends on the shape of the regions and can in principle
be computed explicitly.
\smallskip

\noindent{\bf 5.}
There has been much work done on the `regularized' determinant
of the {\it continuous} Laplacian, and the asymptotic distribution of
its eigenvalues \cite{Kac,MS,OPS}. 
We have not attempted here to make any connection
between these two subjects, although there is a lot of evidence for
a connection, see e.g. \cite{OPS,Dup}.
\smallskip

\noindent{\bf 6.}
As noted above, Temperley \cite{Temp} gave a bijection between 
the set of spanning trees of a 
subgraph of $\Z^2$ and the set of domino tilings of a related polyomino. 
The corollary follows from the theorem by applying this
bijection (see section \ref{defs} for the definition): the graph $H_\eps$ gives
rise to a polyomino $P_\eps$ of area $A_\eps=\eps^2(4N(H_\eps)-B(H_\eps)-4)$ and perimeter
$\mbox{Perim}_\eps=\eps(2B(H_\eps)+4)$. 
Plugging these values into (\ref{main2form}) gives
the formula in the corollary.
\smallskip

\noindent{\bf 7.}
The function $ERR(\eps)$ is unknown although it seems possible that it could
be computed using Toeplitz determinants, as in \cite{MW}.
\bigskip

Part of the motivation for proving Theorem \ref{main2}
is to validate a certain
heuristic, which attempts to explain how the 
presence of the boundary affects the long-range structure of a random tiling.
In particular it attempts to explain how the boundary affects the densities of 
local configurations far from the boundary \cite{Dest}. 
We call this heuristic the `phason strain' principle.

The heuristic is as follows: 
the boundary causes the {\it average height function}
of a tiling (see definition in section \ref{ahf})
to deviate slightly from its entropy-maximizing value of $0$.
At a point in the region where the average height function has nonzero slope,
the ``local'' entropy there is smaller than the maximal possible entropy,
by an amount proportional to the square of the gradient of the 
average height function.
The system behaves in such a way as to 
maximize the total entropy subject to the given
boundary values of the height function, and the resulting 
average height function is the function which minimizes the (integral
of) the square of its gradient. That is, the average height function is
harmonic. This ``explains'' the terms $c_2(\eps)\log\frac1\eps+c_3(U)$ in Theorem \ref{main2}.

Unfortunately the constant $\frac\pi{48}$ appearing in Theorem
\ref{main2} is different from the expected value of the
local entropy as derived in \cite{CKP} (where a rigorous version of the
phason strain principle is proved in a different context): 
in \cite{CKP} the entropy as a function of slope is shown to have
the expansion
$$ent(s,t)=ent(0,0)-\frac{\pi}{16}(s^2+t^2)+ O(\mbox{terms of order $\geq3$}),$$
and where $(s,t)$ are the partial derivatives of the height function.
We may conclude from this discrepancy that the computation in \cite{CKP}
can not be refined to obtain asymptotics of the same
precision as Theorem \ref{main2} above (when applied to the present
case, \cite{CKP} only gives the leading term in (\ref{main2form})).
In particular the phason strain principle can not be considered valid
in this context.
\medskip

The techniques used to prove Theorem \ref{main2} can be applied
to the uniform spanning tree model as well. Indeed, as mentioned above,
there is a close connection between the spanning tree
process on $\Z^2$ and the domino tiling model \cite{Temp,BP,KPW}. 
Many properties of spanning trees on $\Z^2$
translate into computable properties of dominos.
We study here one particular property of a uniform spanning tree:
the distribution of the (unique) arc between two fixed points.
The relevant question about tilings is to count the number of tilings of a region 
with a hole (single square removed). 
To estimate this number, we use the technique of Theorem \ref{main2}:
we cut the region apart up to the hole, and then sew it up again
in such a way as to remove the hole. 
In this way we prove the well-known conjecture that the expected number of points
on the tree branch
within distance $N$ from the origin grows like $N^{5/4}$.
In fact we prove more:
\begin{thm}\label{lerwthm}
On the uniform spanning tree process on $\N\times\Z$, the 
expected number of vertices on the branch from $(0,0)$ to $\infty$
which lie within distance $N$ of the origin is $N^{5/4+o(1)}$. 
For $x>0$ the probability of a vertex $(x,y)=re^{i\theta}$ to
be on the branch from $(0,0)$ to $\infty$
is $$r^{-3/4(1+f(r))}\cos(\theta)^{1/4}(1+o(1))$$
where $f(r)$ is $o(1)$ as $r\to\infty$.
\end{thm}
The branch in a uniform spanning tree
has the same distribution as the loop-erased random walk (LERW), see
\cite{Pem}. So this proves that the growth exponent of the loop-erased
random walk is $\frac54$.
This value of the exponent $d$ has been conjectured by physicists for some time
\cite{Gutt, Maj}, using arguments based on conformal field theory and the assumption 
of conformal invariance of the ``scaling limit'' of the walk.
Lawler \cite{Law} had previously given the bounds $1<d\leq \frac43$.
\medskip

Here is an outline of the paper. 
Section \ref{defs} gives the definitions and background.
Most of the background comes from \cite{Kenyon.prob} and
\cite{Kenyon.ci}: local
properties of dominos can be found in \cite{Kenyon.prob}, and the conformal
properties can be found in \cite{Kenyon.ci}. 
In Section \ref{cutting}, we state two lemmas and use them
to prove Theorem \ref{main2}. 
Specifically, the theorem is proved by cutting up a
rectilinear polygon into rectangles and using the known
formula for the number of tilings of a rectangle. 
Lemma \ref{nextdom} determines how the number of tilings changes
as you are making a single cut, and Lemma \ref{changeinDE} relates this change
to the change in Dirichlet energy of the average height function.
The next two sections are devoted to the proofs of the lemmas. 
Section \ref{rectsect} recalls the formula for the number of 
tilings of a rectangle and proves the formula of Theorem \ref{main2}
in this special case. 
Section \ref{LERW} discusses the connection of domino tilings to 
loop-erased random walk and proves Theorem \ref{lerwthm}.

\medskip
We kindly acknowledge Oded Schramm, Wendelin Werner and Bertrand Duplantier
for helpful discussions, and thank Oded Schramm and
Russell Lyons for proofreading.

\section{Definitions and Background}\label{defs}
\subsection{Temperleyan polyominoes}
By the {\bf grid $\cal G$} we mean the graph whose vertices are $2\Z^2$ and edges
join all pairs of vertices at distance $2$.
A {\bf lattice square} is a face of $\cal G$.
A {\bf simply-connected subgraph} of $\cal G$ is a set of vertices and edges
of the grid which is the $1$-skeleton of a simply connected union
of (closures of) lattice squares. 

\begin{figure}[htbp]
\PSbox{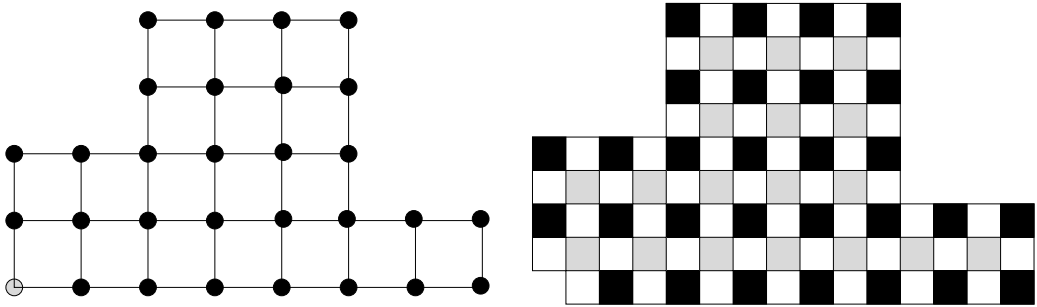}{0in}{3in}
\caption{\label{W.P}The graph $H$ and polyomino $P(H)$.
The squares of $B_0$ are black, those of $B_1$ are in grey.
The vertex $b$ of $H$ is in grey, and corresponds to the missing corner of $P$.}
\end{figure}

Let $H$ be a finite simply-connected subgraph of $\cal G$ and $b\in H$ a fixed
vertex adjacent to the outer face of $H$.
We associate to $H$ a new graph $P'=P'(H)=P'(H,b)$ as follows.
There is a vertex of $P'$ for each vertex, edge and face of $H$,
except for the outer face and the vertex $b$.
Two vertices $u_1,u_2$ of $P'$ are connected by an edge in two cases:
$u_1,u_2$ come from an edge $e$ and a vertex $v$ of $H$ (and 
$v$ is on the edge $e$),
or $u_1,u_2$ come from a face $f$ and edge $e$ of $H$ (and $e$ 
is part of the boundary of $f$). In
other words, $P'(H)$ is the ``superposition" of H and its planar dual $H'$, 
except that we discard vertex $b$ of $H$ and the outer vertex of 
$H'$.  Let $P(H)$ denote the polyomino in $\frac12\cal G$
whose dual (not including the vertex for the outer
face) is $P'(H)$.  See Figure \ref{W.P} for an example.

Except for the outer face, the faces of $P$ are squares, and come in four types,
$B_0,B_1,W_0,W_1$. The squares in $B_0$ are those coming from vertices
of $H$; the squares in $B_1$ are those coming from faces of $H$. The squares
in $W_0$ (respectively $W_1$) are those coming from horizontal (resp. vertical)
edges of $H$. The `B' and `W' stand for `black' and `white' coming
from the checkerboard coloring of $P$. See Figure \ref{W.P}.
We assign colors $B_0,B_1,W_0,W_1$ to vertices
of $P'$ corresponding to the colors of the faces of $P$.

Temperley \cite{Temp} showed that there is a bijection between spanning trees
of $H$ and perfect matchings of $P'(H)$ (a {\bf perfect matching},
or dimer covering, is a set of edges such that every vertex
is contained in a unique edge). 
Perfect matchings of $P'$ correspond to {\bf domino tilings}
(tilings with $1\times 2$ and $2\times 1$ rectangles) of $P(H)$.
A polyomino (a union of lattice squares bounded by a simple closed curve) 
is said to be {\bf Temperleyan} if it is of the form
$P(H)$ for some simply-connected subgraph $H$ of $\cal G$.
Note that if $H$ is a simply-connected subgraph of $\eps \cal G$ for some $\eps>0$,
then $P(H)$ will be a polyomino in $\frac\eps2\cal G$. 

The lattice square missing in $P$ which corresponds to point $b\in H$ 
is called the {\bf base square} of the polyomino.
For the graph $P'$ dual to $P$, $b$ is the {\bf base vertex} (it is not
a vertex of $P'$). 
By Temperley's bijection, the number of tilings of $P$ is independent
of choice of vertex $b\in H$ as long as it is on the boundary.
If $P$ is a Temperleyan polyomino then by $H=H(P)$ we mean the unique associated
simply connected subgraph of $\cal G$ for which $P=P(H)$.

These definitions also apply to infinite graphs $H$, the only difference being
that in this case we may if we like choose $b=\infty$, which means we do not 
remove any lattice square from $P(H)$.
All infinite graphs we deal with in the sequel have the property that near to infinity
the boundaries are straight, that is, the boundaries have no corners
outside some fixed large radius. This is to avoid certain convergence problems later.
Two important examples of infinite Temperleyan polyominos are the whole plane 
$P(\Z^2)$ and the half plane $P(\N\times \Z)$. Both of these have $b=\infty$.

\subsection{Conformal properties}\label{confprops}
The results of \cite{Kenyon.ci} apply to Temperleyan polyominoes.
Let $P$ be a Temperleyan polyomino with dual graph $P'$. 
We assign weights to the edges of $P'$ so that a horizontal edge
has weight $1$ if its left vertex 
is white, $-1$ if its left vertex
is black; a vertical edge is weighted $i=\sqrt{-1}$ if its lower vertex is white,
$-i$ if its lower vertex is black. Thus the weights around a white 
vertex are $1,i,-1,-i$ in counterclockwise order starting from the edge leading right; around a black vertex these
weights are $-1,-i,1,i$.
The adjacency matrix  
of the graph $P'$ with these weights is called the {\bf Kasteleyn matrix} $K_P$
of $P$. Its determinant is the square of the number of domino
tilings of $P$ \cite{Kast2}.
The inverse of the Kasteleyn matrix is called the {\bf coupling function}
$C_P(\cdot,\cdot)$ of $P$. The probability of a configuration of dominos 
occurring in a random
tiling is the absolute value of the determinant of a submatrix of
the coupling function matrix \cite{Kenyon.prob}.

The coupling function has a number of important properties which we list here.
\subsubsection{Combinatorial properties of the coupling function}
First, for $v_1,v_2$ two vertices of $P'$, we have
$C_P(v_1,v_2)=C_P(v_2,v_1)$ and if $v_1$ and $v_2$ are both black or both white
then $C_P(v_1,v_2)=0$. Therefore we will always take the first variable of
$C_P$ to be a white vertex and the second variable to be a black vertex.

The coupling function has a concise description in terms of the Green's function on $H$.
Let $G$ be the Green's function on the graph $H$, that is, $G$ is the function on
$H\times H$ which satisfies $\Delta G(x,y)=\delta_x(y)-\delta_b(y)$, where $\Delta$ is the Laplacian
with respect to the second variable (and recall that $b$ is the base vertex). 
Here $\Delta f(v) = 4f(v)-f(v+2)-f(v-2)-f(v+2i)-f(v-2i)$,
except at a boundary vertex, where $\Delta f(v)$ is the degree of $v$ times $f(v)$
minus the sum of the neighboring values. (Here $v\pm 2,v\pm 2i$ refer
to the four neighbors of $v$ in the graph $H$. Note that
these vertices are at distance $2$ in $P'$.) As stated the function $G$ 
is only well-defined up to an additive constant.
We fix the constant by setting the function to be zero when $y=b$.
For $x,x'$ any two vertices of $H$, the function $L(y):=G(x,y)-G(x',y)$ 
satisfies $\Delta L(y)=\delta_x(y)-\delta_{x'}(y)$
and $L(b)=0.$

Let $x$ and $x'$ be adjacent vertices of $H$; let $f,f'$ be the faces of $H$ adjacent 
to the edge $xx'$, with $f'$ on the left as the edge is traversed from $x$ to $x'$.
Let $\hat L$
be the function on the faces of $H$
which is the harmonic conjugate of $L(y)$
in the sense that for any edge $e=v_1v_2$ of $H$ which is not the edge $xx'$ we have
$L(v_2)-L(v_1)=\hat L(f_1)-\hat L(f_2)$, where $f_1$ is the face to the 
left of the edge $e$ (when $e$ is traversed from $v_1$ to $v_2$)
and $f_2$ is the face to the right. If we define 
$\hat L$
to be zero on the outer face then it is uniquely defined and harmonic except at $f$ and $f'$.
Moreover $\Delta\hat L(\cdot)=\delta_f(\cdot)-\delta_{f'}(\cdot)$,
where $\Delta$ is the Laplacian on the dual $H'$ of $H$. 
The function $\hat L(z)$ can also be written $\hat G(f,z)-\hat G(f',z)$,
where $\hat G$ is defined by $\Delta\hat G(f,z)=\delta_f(z)-\delta_o(z)$ and
$\hat G(f,o)=0$, $o$ referring to the outer face. 

We now have the following description of the coupling function in terms of these Green's
functions.
Suppose $v_1\in W_0$. Then 
\begin{equation}\label{couplgreen1}
C_P(v_1,v_2)= \left\{\begin{array}{ll} G(v_1+1,v_2)-G(v_1-1,v_2)&\mbox{if }v_2\in B_0\\
i(\hat G(v_1+i,v_2)-\hat G(v_1-i,v_2))&\mbox{if }v_2\in B_1.
\end{array}\right.
\end{equation}
Note that when $v_2\in B_0$, 
$C_P(v_1,v_2)= G(v_1+1,v_2)-G(v_1-1,v_2)$ makes sense since both $v_1\pm 1$ and
$v_2$ are in $B_0$ (hence vertices of $H$). 
When $v_2\in B_1$ rather, $C_P(v_1,v_2)=
i(\hat G(v_1+i,v_2)-\hat G(v_1-i,v_2),$ makes sense since $v_1\pm i$ and $v_2$
are in $B_1$ (faces of $H$). 
Similarly, if $v_1\in W_1$ we have 
\begin{equation}\label{couplgreen2}
C_P(v_1,v_2)= \left\{\begin{array}{ll}\hat G(v_1+1,v_2)-\hat G(v_1-1,v_2)&\mbox{if }
v_2\in B_1\\
-i(G(v_1+i,v_2)-G(v_1-i,v_2))&\mbox{if }v_2\in B_0.
\end{array}\right.
\end{equation}

This description of the coupling function follows from the fact that
$K^*K$ is the Laplacian on $H$, and also acts as the Laplacian on the dual of $H$,
see \cite{Kenyon.ci}.

An important case of the above formula for $C_P$ is when $v_1$ corresponds to 
an edge on the boundary of $H$, so that one of $v_1\pm1$ or $v_1\pm i$ 
corresponds to the outer face of $H$.
Suppose for example that $v_1+1$ is the outer face of $P'$;
then $\delta_{v_1+1}(v_2)=0$ by definition (for $v_2$ a vertex of $B_1$,
that is, a face of $H$ which is not the outer face)
and so the term $G(v_1+1,v_2)$ can be ignored in the above formula.

\subsubsection{Asymptotic properties}
Let $U$ be a rectilinear polygon in $\C$.
Fix a base point $b_0\in\partial U$.
Let $\{P_\eps\}_{\eps>0}$ be a sequence of Temperleyan polyominoes
$P_\eps\subset\eps\Z^2,$ approximating $U$ as $\eps\to0$ in the following sense.
\label{Pepsdef}
The $P_\eps$ are rectilinear with the same number of corners
as $U$, one corner converging to each corner of $U$.
Furthermore the base points $b_\eps\in P_\eps$ converge to $b_0$.

In \cite{Kenyon.ci} we proved the following. Let $C_\eps=C_{P_\eps}$.
Under the above convergence hypotheses, 
the rescaled coupling functions $\frac1\eps C_\eps$
converge to a pair of complex-valued functions $F_0(v,z)$ and $F_1(v,z)$ 
which are meromorphic in $z$, in the following sense.
If $\{u_\eps\},\{v_\eps\},\{w_\eps\},\{x_\eps\}$ with $u_\eps,v_\eps,
w_\eps,x_\eps\in P_\eps$ are four sequences of 
vertices of type $W_0,W_1,B_0,B_1$ respectively, converging to respectively
$u,v,w,x\in U$,
then
\begin{eqnarray*}
\lim_{\eps\to0}\frac1\eps C_\eps(u_\eps,w_\eps)=\Re F_0(u,w)\\
\lim_{\eps\to0}\frac1\eps C_\eps(u_\eps,x_\eps)=i\Im F_0(u,x)\\
\lim_{\eps\to0}\frac1\eps C_\eps(v_\eps,w_\eps)=\Re F_1(v,w)\\
\lim_{\eps\to0}\frac1\eps C_\eps(v_\eps,x_\eps)=i\Im F_1(v,x).
\end{eqnarray*}
The functions $F_0$ and $F_1$ are defined by the following properties.

\begin{prop} {\bf\cite[Theorem 13]{Kenyon.ci}} \label{F0F1props}
For each fixed $v\in U$ the function $F_0(v,z)$ has a following properties:
\begin{enumerate}\item it is meromorphic as a function of $z\in U$
\item its imaginary part vanishes for $z\in\partial U$, 
\item it has a zero at $z=b_0$, 
\item it has a
a simple pole of residue $\frac1\pi$ at $z=v,$ and no other poles
on $\overline U$.
\end{enumerate}
Similarly, for each fixed $v\in U$ the function $F_1$ has the following
properties:
\begin{enumerate}\item it is meromorphic 
as a function of $z\in U$; \item
its real part vanishes for $z\in\partial U$,
\item  it has a zero at $z=b_0$, 
\item it has a simple pole of residue $\frac1\pi$ at $z=v,$ and no other poles
on $\overline U$.
\end{enumerate}
Furthermore, $F_0$ and $F_1$ are the {\em unique} functions with these 
properties.
\end{prop}

Define the functions $F_+=F_+^U:=F_0+F_1$ and $F_-=F_-^U:=F_0-F_1$. 
These functions are easier to work with since they transform
as nicely under conformal mappings: $F_+(v,z)dv$ is a meromorphic
$1$-form and $F_-(v,z)d\bar v$ is an antimeromorphic $1$-form.

\begin{prop} {\bf \cite[Proposition 15]{Kenyon.ci}}
The function $F_+(v,z)$ is analytic in both variables. The function
$F_-(v,z)$ is analytic in $z$ and anti-analytic in $v$.
If $V$ is another marked region and 
$f\colon V\to U$ a conformal isomorphism sending the base point
of $V$ to the base point of $U$, then we have the 
transformation rules
\begin{eqnarray}\label{transrules1}F_+^V(v,z) = f'(v)F_+^U(f(v),f(z))\\
F_-^V(v,z) = \overline{f'(v)}F_-^U(f(v),f(z)).\label{transrules2}
\end{eqnarray}
\end{prop}

When $U=\C$ with $b_0=\infty$
we have $F_+(v,z)=\frac2{\pi(z-v)}$ and $F_-(v,z)\equiv 0$.
When $U$ is the right half-plane $RHP=\{z|\Re(z)>0\}$ with $b_0=\infty$ we have 
\begin{equation}\label{RHPF}
F_+(v,z)=\frac2{\pi(z-v)},\hskip1in
F_-(v,z)=-\frac2{\pi(z+\overline{v})}.
\end{equation}
This and the above transformation rules
determine $F_+$ and $F_-$ (and hence $F_0,F_1$) on any 
simply-connected region.

\subsection{Average height function}\label{ahf}
\begin{figure}[htbp]
\begin{center}
\PSbox{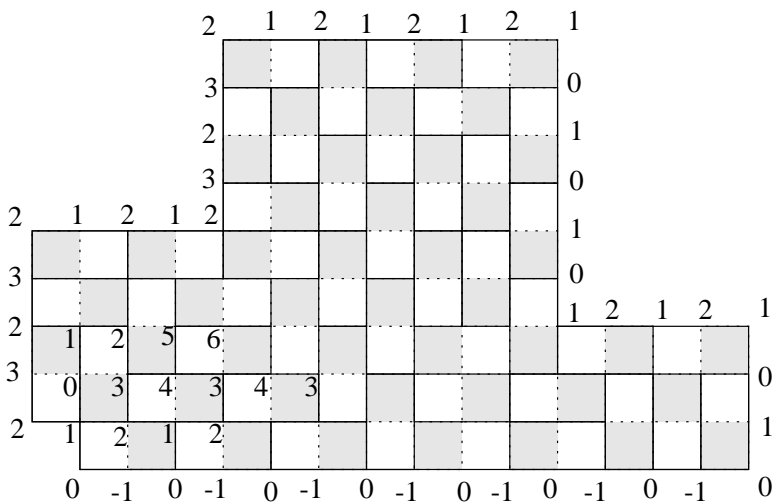}{-5in}{3in}
\end{center}
\caption{\label{heights}Some of the heights in a domino tiling.}
\end{figure}

Recall \cite{Thu} that the height function 
of a domino tiling is an integer-valued
function on the vertices of the dominos; it is well-defined
up to an additive constant. 
After fixing its value at some vertex, it is defined
by the property that on an edge $v_1v_2$ which is not crossed
by a domino, the height difference $h(v_2)-h(v_1)$ is $+1$ if the 
square to the left of the edge $v_1v_2$ (we mean,
the square containing edge $v_1v_2$ and on the left when traversing
$v_1v_2$ from $v_1$ to $v_2$) is black, and $-1$ if
this square is white. See Figure \ref{heights}. Note that
the height function along the boundary is independent of the tiling.

Let $U$ be a rectilinear polygon with base point $b_0$ and
$P_\eps$ a sequence of Temperleyan polyominos
converging to $U$ as in the previous section.
Suppose for simplicity that all the $P_\eps$ contain a fixed vertex $v_0$,
say at a corner of $P_\eps$.
Let $h_\eps$ be the average height function of $P_\eps$,
that is, for any vertex $v\in P_\eps$, $h_\eps(v)$ is the average height
of $v$ over all domino tilings of $P_\eps$, where the height at $v_0$
is taken  to be zero.
The {\bf limiting average height function}  $h$
of a random domino tiling of $U$
is by definition the limit as $\eps\to 0$ of the functions $h_\eps$:
take $x\in U$, with 
$x\not\in\partial U$ and let $x_\eps\in P_\eps$ converge to $x$; then
$$h(x):=\lim_{\eps\to0}h_\eps(x_\eps).$$
For $x\in \partial U$ but not at a corner, $h(x)$ is defined by continuity
from values of $h$ in the interior.

In \cite{Kenyon.ci} we showed that this limit exists and has a simple expression in
terms of the function $F_+$. Let $F_+^*(v,z)=F_+(v,z)-\frac2{\pi(z-v)}$
and let $F_+^*(v)=\lim_{z\to v}F_+^*(v,z)$. 
Then the limiting average height function $h$ on $U$ is
given by the complex line integral $$h(v)=2\Im\int_{v_0}^v F_+^*(u)du.$$
The choice of zero $v_0$ is immaterial since the height
is only defined up to an additive constant anyway.
Note that from (\ref{RHPF}), on $\C$ or on the right half-plane we have
$F_+(v,z)=\frac2{\pi(z-v)}$ so $F_+^*\equiv 0$ and therefore $h(v)$ is constant,
as expected. As another example, the map $z\mapsto\frac{1+z}{1-z}$ maps the unit
disk to the right half-plane, mapping $1$ to $\infty$; therefore using (\ref{transrules1}),
on the unit disk with basepoint $z=1$ we have $F_+(v,z)=\frac{2(1-z)}{\pi(1-v)(z-v)}$
and $F_+^*(z)=\frac{-2}{\pi(1-z)}$, 
so that $h(v)=-\frac4\pi\Im\int_{v_0}^v\frac{du}{1-u}.$

Note that in general $h(v)$ is harmonic, being the imaginary part of an
analytic function.

For $v$ on the outer boundary of $U$,
$h(v)-h(v_0)$ is the total 
turning (in radians times $2/\pi$) of the boundary tangent on the path 
counterclockwise (cclw) around the boundary from $v_0$ to $v$.
However when passing the basepoint $b_0$ the height drops by $4$.
Thus a full cclw turn contributes $+4-4=0$ to $h$.
The function $h$ is constant on the straight edges, and 
there is a discontinuity at each corner of $U$, where $h$ changes by $\pm1$
according to whether the corner is a left turn or right turn
(if the basepoint $b_0$ is at a corner then the change would be $-3$ or $-5$ accordingly).
These boundary values can be understood in a sense
using Figure \ref{heights}: the polyomino
in this figure can be thought of as approximating a rectilinear octagon $U$,
where the height on the lower boundary of $U$ is $-1/2$ (the average
of $-1$ and $0$, the alternating heights on $P_\eps$), the height on the right-most
boundary is $1/2$ (the average of $0$ and $1$) and so on. 
Since a harmonic function is determined by its boundary values
this provides a simple description of $h$
in terms of the turning of the boundary tangent.

\subsection{Dirichlet energy}\label{deltadependence}
Recall that the Dirichlet energy of a harmonic function $h$ 
on a region $U$ is given by 
$$E(h)=\int\!\!\!\int_U|\nabla h|^2\,dx\,dy=\oint_{\p U} h\,dg$$
where $g$ is a harmonic conjugate of $h$, that is, $g$
is a harmonic function so that $h+ig$ is locally an analytic function of $x+iy$.

We will be interested in the Dirichlet energy of
harmonic functions which are the limiting
average height functions on rectilinear polygons
$U$. In particular their boundary values have a finite number of
jump discontinuities (at the corners); unfortunately 
in such a case the Dirichlet energy is infinite. 
To avoid this difficulty, for each sufficiently small $\delta>0$ 
we define the {\bf $\delta$-normalized Dirichlet energy} $E_\delta(h)$
as follows.
Remove a $\delta$-neighborhood of
each $x\in\partial U$ for which the harmonic function $h$ has a jump
discontinuity. Let $U'$ be the region $U$ without these
neighborhoods. The $\delta$-normalized 
energy $E_\delta(h)$ is simply the integral
of $|\nabla h|^2$ over $U'$.

If $U$ is unbounded and if 
$h$ has a jump discontinuity at $\infty$, 
we remove the neighborhood of $\infty$ consisting of points
$|z|>\frac1\delta$, and compute the energy on the remaining region as before. 

Here we will illustrate with an example.
Let $h$ be the bounded harmonic function on the upper half plane
which has value $0$ on the $x$-axis to the right of the origin 
and $1$ on the $x$-axis to the left of the origin.
Then $h(z)=\frac1\pi\Im\log(z)$. The harmonic conjugate to $h$ is $g(z)=
-\frac1\pi\Re\log(z).$ The normalized Dirichlet energy is the integral
over $\partial U'$ of $h\,dg$, which can be broken into four parts:
the integral from $-\delta^{-1}$ to $-\delta$, the integral around the half
circle of radius $\delta$, the integral from $\delta$ to $\delta^{-1}$,
and the integral around the half-circle of radius $\delta^{-1}$.
The third of these integrals is zero since $h$ is zero on the positive $x$-axis.
The second and fourth are zero since $dg$ is zero on circles about the origin.
So the only contribution is from the first integral which gives
$$E_\delta(h)=\int_{-1/\delta}^{-\delta}1dg = g(-\delta)-g(-1/\delta)=
\frac2\pi\log\frac1\delta.$$

In a more general situation $dg$ will not vanish along the 
boundaries of the $\delta$-neighborhoods
of the discontinuities, but $dg$ will still be $O(\delta)$ 
there, as we now show: 
Let $h$ be the average height function on a rectilinear
polygon $U$.  Take a convex corner of $U$,
translate and rotate $U$ so that the corner is at the origin
and is bounded by the positive axes. 
Without loss of generality (after a linear scale)
we suppose that near the origin 
$h$ is $0$ on the $x$-axis and $-1$ on the $y$-axis.
Then $h$ is the imaginary part of an analytic function $\tilde h(z)$ on $U$
whose expansion at $0$ is of the form
$$-\frac2\pi\log(\alpha_1 z+\alpha_2 z^2+\ldots)=-\frac2\pi\log(z)+\beta_0+
\beta_1 z+\beta_2 z^2+\ldots.$$
In particular when $z=\delta e^{i\theta}$, 
we have 
$$\frac{dg(z)}{d\theta}=
\frac2\pi\Re(\beta_1\delta ie^{i\theta}+O(\delta^2))=O(\delta).$$
A similar argument works at a non-convex corner.

Since $h$ has a standard form near each of its
jump singularities, the $\delta$-normalized Dirichlet energy
has a very simple dependence on $\delta$.
Recall that at a convex corner the height function changes by $+1$
when moving cclw around the boundary. So the height function
is of the form of that of the above example.
If we change $\delta$ to a smaller $\delta'$, the change in
energy at that corner is 
$$\int_{i\delta}^{i\delta'}-dg =-\frac2\pi\log(\delta')+\frac2\pi\log\delta+
O(\delta).$$ 
Thus the dependence on $\delta$ at a convex corner is 
$\frac2\pi\log\frac1\delta+O(\delta)$.

Similarly we can do the calculation at a concave corner. Move and 
rotate the corner so that it is bounded by the positive $x$-axis and the
negative $y$-axis. Up to an additive constant $h$ is $0$ on the $x$-axis
near the origin and $1$ on the negative $y$-axis. So $h$
is the imaginary part of an analytic function 
$\tilde h(z)=\frac2{3\pi}\log z+\alpha_0+\alpha_1z+\alpha_2z^2+\ldots.$
Now if $\delta$ changes to the smaller $\delta'$, the change in
energy is $$\int_{-i\delta }^{-i\delta'}dg=-\frac2{3\pi}\log(\delta')
+\frac2{3\pi}\log\delta+O(\delta).$$
Thus the dependence on $\delta$ at a concave corner is 
$\frac2{3\pi}\log\frac1\delta+O(\delta)$.

If a corner contains the base point, then if it is a convex corner
the height changes by $-3$ rather than $+1$. 
the dependence on $\delta$ is therefore $9$ times that for a normal
convex corner, or $\frac{18}\pi\log\frac1\delta$.
So the fact that 
the corner contains the base point adds $\frac{16}\pi\log\frac1\delta$
to its ``local energy". If the corner is concave
the height changes by $5$ rather than $1$, and so the 
local energy 
is $\frac{50}{3\pi}\log\frac1\delta$ rather than $\frac2{3\pi}\log\frac1\delta$,
which is also an addition of $\frac{16}\pi\log\frac1\delta$.
If the base point occurs along an edge, the height change is $4$ and
the energy associated is again $\frac{16}\pi\log\frac1\delta$.

Now from these calculations, 
for any rectilinear region $U$ we can immediately
compute the dependence of the $\delta$-energy on $\delta$,
in terms of the number of vertices. A (simply-connected) region with $V$ vertices
has $(V-4)/2$ concave vertices and $(V+4)/2$ convex vertices,
one of which we may take to be the base vertex.
So the dependence on $\delta$ of its $\delta$-energy is 
$$O(\delta)+
\left(\frac2{3\pi}\left(\frac{V-4}2\right)+\frac2\pi\left(\frac{V+4}2\right)+
\frac{16}\pi
\right)\log\frac1\delta=$$
$$=\left(\frac{4(V-4)}{3\pi}  +\frac{24}\pi\right)\log\frac1\delta + 
O(\delta).$$

Note that $-\pi/48$ times the above $\delta$-energy gives
the logarithmic terms in Theorem \ref{main2} (if we replace $\delta$
with $\eps$).

\section{Cutting a lattice region into rectangles}\label{cutting}
Here we prove Theorem \ref{main2}. 
Let $U$ be a rectilinear polygon with basepoint $b_0\in\partial U$.
Let $\gamma_0=\gamma_0(t)$ be a straight (horizontal or vertical)
unit speed path in $U$ from $\partial U$ to $\partial U$, which avoids $b_0$
and does not touch $\partial U$ except at its endpoints.

Let $P_\eps\subset\eps\Z^2$ be a 
Temperleyan polyomino approximating $U$ as described in section \ref{Pepsdef}.
Let $\gamma_\eps$ be a strip of width $\eps$ of lattice squares of
$P_\eps$, lying within $O(\eps)$ of $\gamma_0$ and
traversing $P_\eps$ from the boundary to the boundary,
avoiding the base square of $P_\eps$. Furthermore we require that $\gamma_\eps$
contain no square in $B_0$ (that is, contains 
only white squares and squares in $B_1$), and
if either extremity of $\gamma_0$ is at a concave
corner of $U$ then the corresponding extremity or extremities of $\gamma_\eps$
are at the corresponding corners of $P_\eps$.

Because of the boundary
conditions on $P_\eps$, $\gamma_\eps$ has length which is an odd multiple of
$\eps$. If we remove $\gamma_\eps$ from $P_\eps$, then what remains is
a union of two disjoint polyominos $P_1$ and $P_2$. 
Let $P_1$ be the polyomino which contains
the base square of $P_\eps$; then $P_1$ is a Temperleyan polyomino.
The other polyomino $P_2$ will become a Temperleyan polyomino if we remove a single
square $s$ of type $B_0$ in $P_2$ adjacent to one of the endpoints of $\gamma_\eps$. 

Note that the union $\gamma_\eps\cup\{s\}$ has
a unique domino tiling. The number of tilings of $P_\eps$ equals the product of
the number of tilings of $P_1$ and the number of tilings of $P_2$, divided by
the probability $\Pr(\gamma_\eps\cup\{s\})$ that the tiling of $\gamma_\eps\cup\{s\}$
occurs in a uniform tiling of $P_\eps$. 

We can repeat this procedure on $P_1$ and $P_2$, cutting them apart into
simpler and simpler pieces until we arrive at a collection of Temperleyan rectangles
(a Temperleyan rectangle is an odd-by-odd rectangle with corners in $B_0$ and
one corner square removed).
Temperley \cite{Temp} provides us with a formula for the exact number
of tilings of a Temperleyan rectangle (see Proposition \ref{rectform}). 
Working by induction, to compute the number of tilings of $P_\eps$ it suffices
to be able to compute the probability of finding, in a random tiling of $P_\eps$,
a tiling of $\gamma_\eps\cup\{s\}$.
We cannot compute this probability exactly but we 
can approximate it sufficiently closely (Lemma \ref{nextdom}).

The region $\gamma_\eps\cup\{s\}$ has a unique tiling which is a chain of dominos.
Starting from the boundary of $P_\eps$, let $a_1,a_2,\ldots,a_N$
be the set of consecutive dominos making up the tiling of $\gamma_\eps\cup\{s\}$.
The $a_i$ are laid 
end-to-end except for $a_N$ which is perpendicular to the others.
The probability of all the $a_i$ being present in a tiling of $P_\eps$
is a product, as $j$ runs from $1$ to $N$,
of the probability that $a_j$ is present, given that $a_1,\dots,a_{j-1}$ are
already present: 
\begin{equation}\label{Pprod}
\Pr(a_1,\ldots,a_N)=\prod_{j=1}^N\Pr(a_j~|~a_1,\ldots,a_{j-1}).
\end{equation}

Suppose that $a_1,\ldots,a_{j-1}$ are present already.
These dominos form a strip running from the boundary of $P_\eps$
to a point in the interior of $P_\eps$. The region 
$P_\eps^{(j)}\stackrel{def}{=}P_\eps\setminus\{a_1,\ldots,a_{j-1}\}$
is again a Temperleyan polyomino, by our hypothesis that the $a_i$
contain only black vertices of type $B_1$.
So computing $\Pr(a_j~|~a_1,\ldots,a_{j-1})$ is a matter of computing the
probability of $a_j$ in a random tiling of this region $P_\eps^{(j)}$.

Let $U_j$ be the region $U$ with a slit cut out along the segment
$\gamma_0([0,2(j-1)\eps])$, and translated by $-2(j-1)\eps$
so that the tip of the cut is at the origin. Then $U_0=U$ (up to translation)
and $U_N$ is a union of two rectilinear polygons. 

We may suppose (after applying a rotation if necessary) that the path $\gamma_0$
is horizontal and goes from left to right.
For $0<j<N$ let $f_j$ be the unique conformal isomorphism sending the
right half-plane $\{z: \Re(z)>0\}$ to $U_j$ which sends
$0$ to $0$ (end of the cut), $\infty$ to the base point $b_0$,
and has expansion $f_j(z)=z^2+O(z^3)$ at the origin
(we do not define $f_0$ or $f_N$). 
\label{Fsection}

\begin{lemma} \label{nextdom} 
Let $F_1^{(j)}$ be the function $F_1$ (coupling function limit) on the region $U_j$.
We have 
$$\frac{\Pr(a_j|a_1,\ldots,a_{j-1})}{\sqrt{2}-1}=
1+\left(\frac{\pi}{\sqrt{2}}\eps F_1^{(j)}(\eps,2\eps) -1\right)+err(j)$$
where $err(j)$ is $o(\max(\frac1{j},\frac1{N-j}))$.
The term in brackets on the right side is $O(\eps)$ when $j$ is not close to $0$
or $N$.
Let $\xi>0$ be a small constant
and $K=K(\eps):=\xi/\eps$. Then this can be written
\begin{equation}\label{3poss}
\frac{\Pr(a_j|a_1,\ldots,a_{j-1})}{\sqrt{2}-1}=\left\{
\begin{array}{ll}
\displaystyle
1+\frac{C_0}j+\mbox{err}_1(j)+O(\eps)&\mbox{if }j\leq K\\
\displaystyle
1+ \frac{\eps}{6}S\sqrt{f_j}(0) +\frac1\xi o(\eps)&\mbox{if }K<j<N-K,\\
\displaystyle
1+\frac{C_1}{N-j}+\mbox{err}_2(N-j)+O(\eps)&\mbox{if }N-K\leq j
\end{array}\right.
\end{equation}
where $S\sqrt{f_j}$ is the Schwarzian derivative of $\sqrt{f_j}$ and 
$\mbox{err}_1(x)$ and $\mbox{err}_2(x)$ are $o(\frac1x)$. Here $\mbox{err}_1$ depends only on
whether $\gamma_0$ starts at an edge or at a concave vertex, and $\mbox{err}_2$
depends only on whether $\gamma_0$ ends on an edge or at a vertex.
The constants $C_0,C_1$ are determined as follows.
If $\gamma_0$ starts on an edge, then $C_0=-\frac18$; if $\gamma_0$ starts at a corner
then $C_0=-\frac5{72}$. If $\gamma_0$ ends at an edge then $C_1=-\frac38$;
if $\gamma_0$ ends at a corner then $C_1=-\frac{23}{72}$.
\end{lemma}

As we will see in the proof, the first and third expressions
on the right hand side of (\ref{3poss})
are special cases of the middle expression, except for the error terms.

The probability of Lemma \ref{nextdom} can be
related to the change in normalized Dirichlet energy 
of the limiting average height function of $U_j$:
\begin{lemma}\label{changeinDE} Let $\xi$ and $K$ be as in the previous
lemma.  For $N>j>1$ the difference in the $\delta$-normalized 
Dirichlet energy of the limiting average height function
between $U_j$ and $U_{j-1}$ is 
$E_\delta(h_j)-E_\delta(h_{j-1})=$
\begin{eqnarray*}-\frac{48C_0}{\pi j}+O(\eps)&&\mbox{if }j\leq K\\
-\frac{8\eps}{\pi}S\sqrt{f_j}(z_j) +\frac1\xi o(\eps)&& \mbox{if }K<j<N-K,\\
-\frac{48C_1}{\pi(N-j)}+O(\eps)&& \mbox{if }N-K\leq j
\end{eqnarray*}
where $C_0$ and $C_1$ are defined as in the previous lemma.
When $j=1$ or $j=N$, the difference in energy $E_\delta(h_j)-E_\delta(h_{j-1})$
is a constant depending (for $j=1$) only on whether the cut starts at a corner or edge
and (for $j=N$) on whether the cut ends at a corner or edge.
\end{lemma}

These two lemmas, and a computation of the number of tilings
of a Temperleyan rectangle, give us the main result:
\medskip

\noindent{\bf Proof of Theorem \ref{main2}.}
The proof is by induction on the number of cuts required to cut
$P_\eps$ apart into Temperleyan rectangles.
In the case $P_\eps$ is a Temperleyan rectangle, $V=4$ and
Proposition \ref{epsrectform} below shows that 
(\ref{main2form}) equals the log of the number of tilings, up to an error
$O(\eps)$.

For a general $P_\eps$ as in the statement,
let $\gamma_\eps\cup\{s\}$ be a strip as explained above, cutting $P_\eps$ apart into
two Temperleyan regions $P'$ and $P''$.
We compute the log of the probability of $\gamma_\eps\cup\{s\}$ occurring
in a tiling of $P_\eps$. This probability depends on whether $\gamma_\eps$ starts on an
edge or at a corner, and whether it ends on an edge or at a corner.

Let $a_1,\dots,a_N$ be the chain of dominos of $\gamma_\eps\cup\{s\}$, and let
$P_\eps^{(j)}=P_\eps\cup\{a_1,\dots,a_j\}$.
By Lemma \ref{nextdom} we have 
\begin{eqnarray}
&&\log\frac{\Pr(\gamma_\eps\cup\{s\})}{(\sqrt{2}-1)^N}\label{sums}
=\sum_j\log\frac{\Pr(a_j|a_1,\dots,a_{j-1})}{\sqrt{2}-1}=\\
&=&\nonumber\sum_{j\leq K}\log\left(1+\frac{C_0}j+\mbox{err}_1(j)+O(\eps)\right)
+\sum_{K<j<N-K}\log\left(1+\frac{\eps}6S\sqrt{f_j}(0)+\frac1\xi o(\eps)\right)+\\
&&\nonumber\hskip1in
+\sum_{N-K\leq j}\log\left(1+\frac{C_1}{N-j}+\mbox{err}_2(N-j)+O(\eps)\right)\\
&=&\nonumber\sum_{j\leq K}\frac{C_0}j+\mbox{err}_3(j)+O(\eps)
+\sum_{K<j<N-K}\frac{\eps}6S\sqrt{f_j}(0)+\frac1\xi o(\eps)
+\sum_{N-K\leq j}\frac{C_1}{N-j}+\mbox{err}_4(N-j)+O(\eps).
\end{eqnarray}
In this expression the terms $\mbox{err}_3(j)$ and $\mbox{err}_4(N-j)$ sum to give an error
${ERR}_0(\eps)\log\frac1\eps$ where ${ERR}_0(\eps)$ is $o(1)$.
The remaining terms are equal by Lemma \ref{changeinDE} to $-\pi/48$ times the change
in Dirichlet energy on $\gamma_\eps$, up to the error terms $O(\eps)$ and $\frac1\xi o(\eps)$.
However the terms $O(\eps)$ sum to $KO(\eps)=O(\xi)$ which is negligible.
Similarly the terms $\frac1\xi o(\eps)$ sum to $\frac1{\eps\xi} o(\eps)$ which tends to zero if
$\xi$ tends to zero sufficiently slowly.

Since each tile (except the last) decreases the area by $2$ and increases the perimeter by $4$,
noting that $-2c_0+4c_1=\log(\sqrt{2}-1)$ which cancels the denominator
of (\ref{sums}), the formula (\ref{main2form}) is
correct up to a justification of the term ${ERR}_0(\eps)$.

The error $ERR_0(\eps)$ is the sum of two error terms,
one coming from the beginning of $\gamma_\eps$ and one from the end.
Let
$ERR_1(\eps), ERR_2(\eps)$ respectively be
$\sum_{j\leq K}\mbox{err}_3(j)$ when $\gamma_\eps$ starts at an edge or a 
corner respectively.
Let $ERR_3(\eps),ERR_4(\eps)$ respectively be
$\sum_{j\geq N-K}\mbox{err}_4(N-j)$ when $\gamma_\eps$ ends at
an edge or a corner respectively.  We have
$ERR_1(\eps)+ERR_4(\eps) = ERR_2(\eps)+ERR_3(\eps)$ since when $\gamma_\eps$ is traversed
in either direction the same error occurs (that is, the probability
of $\gamma_\eps\cup \{s\}$ is independent of the position of $s$ as long as it is
of type $B_0$ and on the correct side of $\gamma_\eps$; therefore 
$\gamma_0$ can be traversed in either direction, giving the same probability for
$\gamma_\eps\cup\{s\}$).
Moreover by Proposition \ref{epsrectform}, $ERR_1(\eps)+ERR_3(\eps)=0$, since when we
cut a rectangle apart into two rectangles there is no error (or rather this error, when
multiplied by $\log\frac1\eps$, is still $O(\eps)$). 

Therefore if $\gamma_\eps$ begins at a corner and ends at an edge, or vice versa,
the error $ERR_0$ is $ERR_2-ERR_1$. When $\gamma_\eps$ begins and ends at a corner
the error $ERR_0$ is $2(ERR_2-ERR_1)$. When we cut $P_\eps$ apart to make rectangles,
each concave corner gets cut exactly once, so the total accumulated
errors will be the number of concave corners times a fixed error
$ERR_2(\eps)-ERR_1(\eps)$. Setting ${ERR}(\eps)=ERR_2(\eps)-ERR_1(\eps)$
completes the proof.
\hfill{$\square$}
\bigskip

As a shortcut for computing the log probability of the cut $\gamma_\eps\cup\{s\}$,
from Lemma \ref{nextdom}
the log of (\ref{Pprod}) is $N\log(\sqrt{2}-1)$
plus the sum over $j$ of $\log\left(1+\left(\frac\pi{\sqrt{2}}\eps 
F_1^{(j)}(\eps,2\eps)-1\right)\right)$, 
plus the error term. 
When $\eps$ is small the sum over $j$ can be replaced by an integral
using the variable $t=2\eps j$:
\begin{equation}\label{logprobint}
\Pr(\gamma_\eps\cup\{s\})=N\log(\sqrt{2}-1)+\int_{2\eps}^{2\eps(N-1)}
\left(\frac\pi{\sqrt{2}}\eps F_1^{(t/2\eps)}(\eps,2\eps)-1\right)\cdot\frac{dt}{2\eps}
+{ERR_0}(\eps)\log\frac1\eps+const+o(1).
\end{equation}
Here $ERR_0(\eps)$ depends on whether or not $\gamma_0$ begins or ends at a corner,
but not on $U$.
This form will be useful in section \ref{LERW}.

\section{The probability of the next domino on a cut}
\subsection{The slit plane}
Define the {\bf slit plane} $SP$ to be the plane minus the left half of the $x$-axis:
$SP=\R^2-(-\infty,0]$. For $\eps>0$ define polyomino
$SP_\eps$
to be the (infinite) polyomino obtained from $P(2\eps\Z^2+(0,\eps))$
by removing the lattice
squares centered at $(-k\eps,0)$ for all $k>0$. We assume the lattice square centered at the
origin is of type $W_1$.
The infinite polyomino $SP_\eps$ is Temperleyan, with base point $b$ at infinity
($SP_\eps$ is gotten from the graph $H_\eps$ which is obtained from $2\eps\Z^2+(0,\eps)$
by removing edges 
$(-2k\eps,-\eps)(-2k\eps,\eps)$
for all $k>0$).

For later use, we compute the asymptotic coupling function on the 
slit plane $SP$.
\begin{lemma}\label{slitplane}
The asymptotic coupling function on $SP$ with base point at $\infty$ satisfies
$$F^{SP}_+(v,z) = \frac1{\pi\sqrt{v}}\frac1{(\sqrt{z}-\sqrt{v})}$$
$$F^{SP}_-(v,z) = \frac{-1}{\pi\sqrt{\bar v}}\frac1{(\sqrt{z}+\sqrt{\bar v})}.$$
\end{lemma}
\begin{proof} The map $z\mapsto\sqrt z$ maps $SP$ to the right half-plane $RHP$ and $\infty$
to $\infty$. On $RHP$ we have
$F^{RHP}_+(v,z)=\frac2\pi\frac1{(z-v)}$ and $F^{RHP}_-(v,z)=\frac{-2}\pi\frac1{(z+\bar v)}$.
The result follows using the transformation rules (\ref{transrules1}),
(\ref{transrules2}).
\end{proof}

Let $U,U_j$ be as in section \ref{Fsection}. Let $f=f_j$ be as in that section.
Define $b=b_j$ and $c=c_j$ to be the coefficients in the expansion
\begin{equation}\label{expF}f(z) = z^2+bz^3+cz^4+O(z^5).\end{equation}
Note that $b\in i\R$ and $c\in \R$ since $f$ maps the imaginary axis to the real
axis.
The inverse of $f$ has the expansion
\begin{equation}\label{expFi}f^{-1}(z) = z^{1/2}-\frac b2z+\left(
\frac{5b^2}8-\frac c2\right)z^{3/2}+O(z^2).\end{equation}
The expansion of $\sqrt{f}$ is
\begin{equation}\label{expf}\sqrt{f(z)}=z+\frac b2z^2+
\left(\frac c2-\frac{b^2}8\right)z^3+O(z^4)
\end{equation}
and the Schwarzian derivative of $\sqrt{f}$ at the origin is defined as
\begin{equation}\label{Sf}
S\sqrt{f}(0):=\frac{(\sqrt{f})'''}{(\sqrt{f})'}-\frac32
\left(\frac{(\sqrt{f})''}{(\sqrt{f})'}\right)^2=3c-\frac{9b^2}4.
\end{equation}

\subsection{Proof of Lemma \protect\ref{nextdom}}
Here we prove Lemma \ref{nextdom}.
After a translation we may assume that the right-hand square of the domino $a_{j-1}$
is the square centered at the origin in $\eps\Z^2$.
Then $\Pr(a_j|a_1,\dots,a_{j-1})=C_j(\eps,2\eps)$ where  $C_j$ is the coupling
function on $P_\eps^{(j)}$ translated as above.
We must then approximate $C_j(\eps,2\eps)$. Recall that $\eps$ is a lattice square of type $W_1$
and $2\eps$ is of type $B_1$.

Recall from (\ref{couplgreen2}) 
that for $v\in W_1$, the coupling function $C_j(v,z)$ on $P_\eps^{(j)}$ satisfies
$\Delta  C_j(v,z)=\delta_{v+\eps}(z)-\delta_{v-\eps}(z)$, 
for $z\in H'$ the dual graph of $H=H(P_\eps^{(j)})$,
where $\Delta$ is the Laplacian on $H'$. 
Furthermore $C_j(v,z)$ is zero when $z$ is the outer face of $H$.
In particular $\Delta C_j(\eps,z)=\delta_{2\eps}(z)$, that is,
$C_j(\eps,z)$ is the discrete Green's function on $H'$ centered at $\eps$.
To compute $C_j(\eps,2\eps)$, we will use a theorem of Kesten comparing the values
of a continuous and a discrete harmonic function near the tip of a cut:

\begin{thm}[\cite{Kesten}]\label{Kestensthm} Let $D_{slit}$ be the slit disk
$\{|z|< 1\}\setminus(-1,0]$ and let $\tilde g$ be
a continuous harmonic function on $D_{slit}$ with piecewise continuous boundary
values and boundary values $0$ on the slit. Let
$g_\eps$ be a discrete harmonic function on $\eps\Z^2\cap D_{slit}$ 
with boundary values within $O(\eps)$ of the boundary values of $\tilde g$,
and boundary values $0$ on the slit. 
Then for any $(k,\ell)\in\Z^2$,  
$$g_\eps(k\eps,\ell\eps)=\lambda_{k,\ell}\tilde g(k\eps,\ell\eps) + o(\sqrt{\eps})$$
where the constant $\lambda_{k,\ell}$ only depends on $k$ and $\ell$.
\end{thm}

We cannot apply this theorem directly since in our case
$C_j(\eps,z)$ is not harmonic at the point $z=2\eps$.
However let $C_{SP_\eps}(v,z)$ be the coupling function on the discrete
slit plane $SP_\eps$.
The function $g(z):= C_j(\eps,z)-C_{SP}(\eps,z)$ 
is harmonic for $z\in B_1$ in a neighborhood of the origin in $H'$,
{\em including} the point $z=2\eps$,
since the Laplacians of $C_j$ and $C_{SP_\eps}$ are equal there.
Furthermore $g(z)=0$ (Dirichlet boundary conditions) when $z$ is on the cut $\gamma_0$.
In fact $g(z)$ is discrete harmonic on all of $P_\eps^{(j)}$ except
possibly where the interior of $P_\eps^{(j)}$ meets the negative $x$-axis
somewhere to the left of the slit (but since $C_{SP_\eps}$ admits
an analytic continuation around the origin,
we can define $g(z)$ so as to be harmonic on all of $P_\eps^{(j)}$).

We can therefore use Kesten's theorem applied to $g(z)$
to compute $g(2\eps)=\lambda_{1,0}\tilde g(2\eps)$, where
$\tilde g(z)$ is the continuous harmonic function on $P_\eps^{(j)}$ with the same boundary
values as $g(z)$. It remains to compute $\tilde g(z)$ and $C_{SP_\eps}$.

\begin{lemma}\label{Max}
On $SP_\eps$ with $z\in B_1$, and $z$ not within $O(1)$ of the origin we have
$C_{SP_\eps}(\eps,z)=\tau\eps F_1^{SP}(\eps,z)+o(\eps)=\tau\sqrt{\frac{\eps}{z}}+o(\eps)$, 
for some constant $\tau$.
\end{lemma}

For the proof see the appendix.
Now to compute $\tilde g(z)$, from the lemma it is the (continuous) harmonic with boundary values
$-\tau\eps F_1^{SP}(\eps,z)+o(\eps)$ on $\partial U_j$. However note that the function 
$\eps F_1^{(j)}(v,z)-\eps F_1^{SP}(v,z)$ as a function of $z$
is continuous, harmonic, with boundary values $-\eps F_1^{SP}(v,z)$ on $\partial U_j$.
Multiplying by $\tau$, at $v=\eps$ we must have
$\tilde g(z) = \tau\eps (F_1^{(j)}(\eps,z)-F_1^{SP}(\eps,z))+o(\eps).$

Let $N_L(0)$ be the neighborhood of radius $L$ of the origin, where $L$ is 
chosen small enough so that $N_L(0)\subset P_\eps^{(j)}$. 
More precisely, if $j\leq K$ or $j\geq N-K$ we take $L= \min\{j\eps,(N-j)\eps\}$
and for $K<j<N-K$ take $L=\xi$.

On $\partial N_L(0)$, $\eps F_1^{SP}(\eps,z)$ is $O(\sqrt{\frac{\eps}{L}})$ 
(see Lemma \ref{slitplane}), and
$\eps F_1(\eps,z)$ is comparable to $\eps F_1^{SP}(\eps,z)$ and therefore also
$O(\sqrt{\frac{\eps}{L}})$.   Therefore on $\partial N_L(0)$, 
$\sqrt{\frac{L}{\eps}}(\tau \eps F_1(\eps,z)-\tau \eps F_1^{SP}(\eps,z))$ is bounded
and we have by Kesten's theorem (for the disk of radius $L$)
$$\sqrt{\frac{L}{\eps}}g(2\eps)=\lambda_{1,0}\tau\sqrt{\frac{L}{\eps}}
\left(\eps F_1(\eps,2\eps)-\eps F_1^{SP}(\eps,2\eps)\right) 
+o(\sqrt{\frac{\eps}{L}}),$$
or
$$g(2\eps)=\lambda_{1,0}\tau\left(\eps F_1(\eps,2\eps)-\eps F_1^{SP}(\eps,2\eps)\right) 
+o(\frac{\eps}{L}).$$

Now from Lemma \ref{slitplane} (using $F_1=\frac12(F_+-F_-)$) we have
$\eps F_1^{SP}(\eps,2\eps)=\frac{\sqrt{2}}\pi,$
and plugging in the value of $\lambda_{1,0}\tau$ from
(\ref{lambdaval}) below yields the first result.

Let $f=f_j\colon RHP\to U_j$ be as in the statement.
Let $s(z)=f^{-1}(z)$. 
Then from the transformation rules (\ref{transrules1}), (\ref{transrules2})
and (\ref{RHPF}) we have 
$$F_+^{U}(v,z)=\frac{2s'(v)}{\pi(s(z)-s(v))},$$
$$F_-^{U}(v,z)=-\frac{2\overline{s'(v)}}{\pi(s(z)+\overline{s(v)})}.$$
Consequently
$$\eps F_1(\eps,2\eps)=\frac{\eps}\pi\left( \frac{s'(\eps)}{s(2\eps)-s(\eps)}+
\frac{\overline{s'(\eps)}}{s(2\eps)+\overline{s(\eps)}}\right).$$
Using the expansion (\ref{expFi}) and (\ref{Sf}), and the facts that $b\in i\R$,
$c\in\R$ this reduces to 
$$\frac{\sqrt{2}}\pi + \frac{\sqrt{2}}{6\pi}S\sqrt{f_j}(0)\eps + 
O(\eps^{3/2}).$$
So then 
$$Pr(a_j|a_1,\dots,a_{j-1})=\sqrt2-1+
\lambda_{1,0}\tau\frac{\sqrt{2}}{6\pi}S\sqrt{f_j}(0)\eps+o(\frac{\eps}{\xi}).$$
Plugging in for $\lambda_{1,0}\tau$ gives the center result in (\ref{3poss}).
\medskip

When $j$ is close to $0$ we must replace $\eps$ in 
Theorem \ref{Kestensthm} by $\frac1j$ since $j$ is the combinatorial distance
(number of lattice points in $H$)
to the boundary of the region.  We have
\begin{equation}\label{Pjsmall}
Pr(a_j|a_1,\dots,a_{j-1})=
\sqrt{2}-1+\lambda_{1,0}\tau\frac{\sqrt{2}}{6\pi}S\sqrt{f_j}(0)\eps+
o(\frac1j).\end{equation}
A similar expression holds when $j$ is close to $N$.

Now when $j$ is small or close to $N$ the Schwarzian derivative of
$f_j$ is blowing up in a standard way which is independent of $U$
up to an error $O(1)$:
\begin{lemma}\label{smallj}
When $j$ is small the germ of $f_j$ at the origin is independent
of $U$ (depending only on whether or not the cut starts
at a corner or on an edge) up to an error $O(1)$.
Similarly when $j$ is near $N$ the germ of $f_j$ at the origin
is independent of $U$ up to an error $O(1)$,
only depending on whether the cut ends at an edge or a corner.
\end{lemma}

The proof is in the appendix.
Below when we write $O_z(1)$ we mean an analytic function in $z$
each of whose coefficients is $O(1)$ (as $j\to0$ or $j\to N$). 
In particular $z^3O_z(1)$ is an analytic
function whose $2$-jet vanishes.

Recall that $f_j$ is normalized as in (\ref{expF}).
If $\gamma_0$ starts at an edge of $U$, then when $j$ is small
$f_j$ is approximated by
the map $2\sqrt{\eps jz^2+(\eps j)^2}-2\eps j$ 
which maps $RHP$ to $\{x+iy~|~x>-2\eps j\}-[-2\eps j,0]$.
That is, from Lemma \ref{smallj} we have
$$f_j(z)=2\sqrt{\eps jz^2+(\eps j)^2}-2\eps j+z^3O_z(1)=z^2-\frac{z^4}{4\eps j}
+z^3O_z(1).$$
Therefore $S\sqrt{f_j}(0)= -\frac{3}{4\eps j}+O(1).$
Plugging this into (\ref{Pjsmall}) yields
$$Pr(a_j|a_1,\dots,a_{j-1})=\sqrt2-1-\frac{\sqrt2-1}{8j}+o(\frac1j)+O(\eps)$$
where the $o(\frac1j)$ term is independent of $U$:
the term $o(\frac1j)$ only depends on the error in Kesten's theorem as applied to the
coupling function on $\{x+iy~|~x>-2\eps j\}-[-2\eps j,0]$.

If $\gamma_0$ starts at a corner of $U$, then $f_j$ is
approximated by a map of the form
$$f(z)=C + C'(z+2iB)\sqrt{z-iB}+z^3O_z(1),$$
for constants $C,C'$ and $B$ (this follows from the Schwarz-Christoffel
formula: the derivative of $f$ is a constant 
times $\frac{z}{\sqrt{z-iB}}$ for some real
$B$). The conditions that $f(iB)=-2\eps j$
and $f(z)=z^2+O(z^3)$ determine $C,C',B$ and a short computation gives
$$f_j(z)=z^2-\frac{2i\sqrt3}{9\sqrt{\eps j}}z^3-\frac{z^4}{4\eps j}+O(z^6)+
z^3O_z(1),$$
whence
$$S\sqrt{f_j}(0)=-\frac{5}{12\eps j}+O(1).$$

\begin{figure}[htbp]
\begin{center}
\PSbox{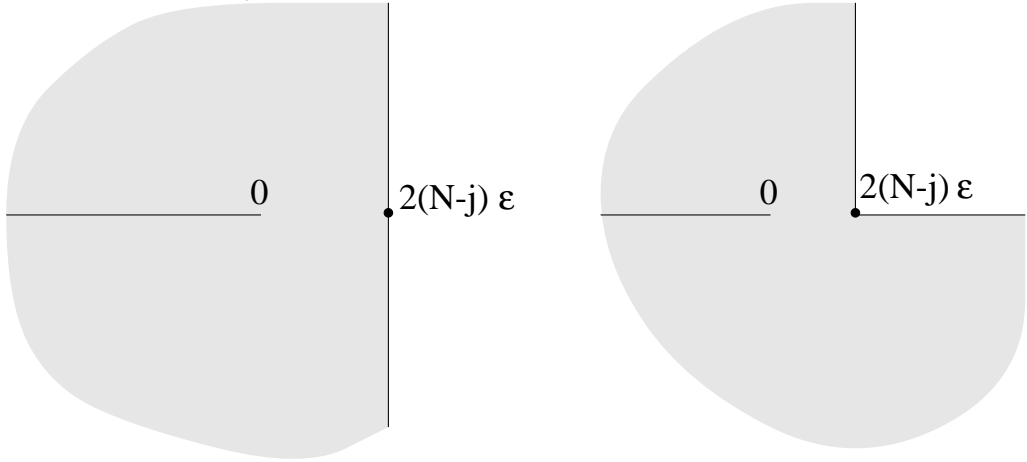}{-5in}{3in}
\end{center}
\caption{\label{elbow}When the cut ends at an edge (a) and at a corner (b).}
\end{figure}

A similar argument holds when $j$ is near to $N$. 
If $\gamma_0$ ends at a straight edge of $U$,
then near the origin $f_j$ is approximated by the map
$$f_j(z)=C-\frac{C'}{\sqrt{z^2+B^2}}+z^3O_z(1),$$
which maps $RHP$ to the region shown in Figure \ref{elbow}a.
Again $C,C',B$ are determined by $f(\infty)=2\eps(N-j)$ and $f(z)=z^2+O(z^3)$,
and a calculation gives $S\sqrt{f_j}(0)=-\frac{9}{4\eps j}+O(1)$.

Lastly, if $\gamma_0$ ends at a corner of $U$, then
near the origin $f_j$ is approximated by a map of the form
$$f_j(z)=C-\frac{C'}{\sqrt{z-iB}(z+2iB)},$$
which maps $RHP$ to the region shown in Figure \ref{elbow}b.
Using $f(z)=z^2+O(z^3)$ and $f(\infty)=2\eps(N-j)$ determines $C,C',B$ to give 
$$f(z)=z^2+\frac{2i\sqrt{3}}{9\sqrt{\eps j}}z^3-\frac{3}{4\eps j}z^4+O(z^5)+z^3
O_z(1),$$
which gives $S\sqrt{f_j}(0)=-\frac{23}{12\eps j}+O(1)$.

This completes the proof.
\hfill{$\square$}
\medskip

To determine $\lambda_{1,0}\tau$, we use Proposition \ref{rectform} below.
When we cut an $m\times n$ rectangle $R$ into two rectangles $R_1$ and $R_2$
with a horizontal cut of length $m$, 
the sum of the logs of number of tilings of $R_1$ and $R_2$,
minus the log of the number of tilings of $R$, which 
is the probability of the cut, is $m\log(\sqrt{2}-1)-\frac12\log m+O(1)$.
On the other hand from the above proof this is
$$m\log(\sqrt{2}-1)+\lambda_{1,0}\tau\frac{\sqrt{2}}{6\pi(\sqrt{2}-1)}\left(
-\frac34-\frac94\right)\log(m)+o(\log m).$$
This gives 
\begin{equation}\label{lambdaval}
\lambda_{1,0}\tau=\frac{(\sqrt{2}-1)\pi}{\sqrt{2}}.\end{equation}

\section{The change in the Dirichlet energy}\label{DEsect}
We prove here Lemma \ref{changeinDE}.
The computation is in two steps. We first show that the change in the 
$\delta$-normalized Dirichlet
energy only depends (up to an error $o(\eps)$) on the $4$-jet of $f$ at the origin,
that is, only depends on $b$ and $c$ of (\ref{expF}).
Then we do an explicit computation of the Dirichlet energy 
for a particular $2$-parameter family of regions, whose corresponding
maps $f$ have $4$-jets covering every possible value of $(b,c)$.

\subsection{Dependence on $4$-jet}
Since in section \ref{deltadependence} we computed the dependence on $\delta$ 
of $E_\delta$, we can assume without loss of generality in this section that
$\delta$ is small compared to $\eps$.

Let $h_j$ denote the limiting average height function for $P_j$ and $h_{j+1}$
be the limiting average height function for $P_{j+1}$.
\begin{lemma}\label{hcircf} Up to an additive constant we have
$h_j\circ f_j= \frac2\pi\Im\log f_j'(z)$. In particular
$h_j=h_j\circ f_j\circ f_j^{-1}=-\frac2\pi\Im\log (f_j^{-1})'(z).$
\end{lemma} 
The proof of the first statement 
follows from the definition of $h_j$ in the last paragraph of section 
\ref{ahf}. The second statement follows trivially from the first.

Let $g_j$ be a harmonic conjugate on $U_j$ of $h_j$,
so that $\tilde h_j:=h_j+ig_j$ is analytic on $U_j$.
Similarly let $g_{j+1}$ be a harmonic conjugate for $h_{j+1}$ on $U_{j+1}$.

Since the boundary of $U_j$ is a subset of the boundary of $U_{j+1}$,
the change in normalized Dirichlet energy can be written
(recall that $U_j'$ is the region $U_j$ minus a $\delta$-neighborhood 
of its vertices)
\begin{equation}\label{U-UU}
\oint_{\p U_{j+1}'} h_{j+1}\,dg_{j+1}-\oint_{\p U_j'}h_j\,dg_j
=\oint_{\p U_j'}\left(h_{j+1}\,dg_{j+1}-h_j\,dg_j\right)+
\oint_{X}h_{j+1}\,dg_{j+1}\end{equation}
\begin{equation}\label{3ints}
= \oint_{\p U_j'}(h_{j+1}-h_j)dg_{j+1}-\oint_{\p U_j'}h_jd(g_{j+1}-g_j) + 
\oint_{X}h_{j+1}dg_{j+1}\end{equation}
where $X=\p U_{j+1}'-\p U_j'$ 
is the path which consisting of the four pieces:
$\partial N_\delta(0)\cup[-\delta,2\eps+\delta]\cup 
\partial N_\delta(2\eps)\cup[-\delta,2\eps-\delta]^*$ 
where the superscript $^*$ is a reminder that the second
segment is traced in the reverse direction.

We'll show that each of the three integrals of (\ref{3ints}) depends
(up to controlled errors) only on the $4$-jets of $f_j$ and $f_{j+1}$.

The first integral in (\ref{3ints}) can be estimated as follows. On
the boundary of $U_j$ we have $h_{j+1}=h_j$, 
and for each corner $c$ of $U_{j+1}$
we have $dg_{j+1}=O(\delta)$ on $\partial N_\delta(c)$
(see section \ref{deltadependence}).
It remains to consider the boundary $\p N_\delta(0)$, where $0$
is the tip of the cut in $U_j$ (which is not a corner
of $U_{j+1}$). The $x$-axis divides $\p N_\delta(0)$ 
into two parts.  Since $g_{j+1}$ is smooth on each half, when $\delta\to0$
the integral of $(h_{j+1}-h_j)dg_{j+1}$ on each half tends to zero.
In conclusion the first integral in (\ref{3ints}) tends to zero with $\delta$.

To compute the second integral in (\ref{3ints}), 
we show that we can replace the path of integration by
a smaller path of radius $O(\eps)$; on this new path we will
be able to replace $dg_j$ and $dg_{j+1}$ with their $4$-jets.

Now $h_{j+1}-h_j$ is the harmonic function which 
(after choosing an appropriate base point)
is zero on $\p U_j$, and $1$ and $-1$ on the upper and lower side
of $[0,2\eps]$, respectively. 
On $\p U_j$ we therefore have 
$$g_{j+1}-g_j=\Im(\tilde h_{j+1}-\tilde h_j)= -i(\tilde h_{j+1}-\tilde h_j)$$ 
since the real part of this function vanishes. 

The second integral in (\ref{3ints}) can then be written
(using $h_j=-\frac2\pi\Im\log (f_j^{-1})'(z)$ from Lemma \ref{hcircf})
\begin{equation}\label{int2}
\oint_{\p U'}h_j\,d(g_{j+1}-g_j)=-\frac2\pi\Im\oint_{\p U'}\log 
(f_j^{-1})'(z)(-i)\left(\tilde h_{j+1}-\tilde h_j\right)'dz,
\end{equation}
where the $'$ refers to the $z$-derivative.
This integral is the imaginary part of a contour integral. 
The integrand is analytic on $U'-[0,2\eps]$;
in particular it is analytic on $U'$ minus a neighborhood of the
origin of radius $3\eps$.
Therefore we can replace the path of integration by a path which winds
around the boundary of the slit disk 
$D_{slit} = \{z\in U': |z|<3\eps\}.$

Now on $D_{slit}$ we have (see (\ref{expFi}))
$$\log(f_j^{-1})'(z)=\log(\frac1{2\sqrt{z}}-\frac{b}2+\gamma\sqrt{z}+zO_z(1))=
\log((f_{j,\{4\}}^{-1})')(z)+z^{3/2}O_z(1)$$
where $f_{j,\{4\}}$ is the $4$-jet of $f_j$ at the origin ($\gamma$ is a constant
depending on $b,c$ only).
Similarly 
\begin{eqnarray*}\log(f_{j+1}^{-1})'(z)&=&\log(\frac1{2\sqrt{z-2\eps}}-\frac{b_{j+1}}2+
\gamma_{j+1}\sqrt{z-2\eps}+O(z-2\eps))\\
&=&
\log((f_{j+1,\{4\}}^{-1})')(z)+(z-2\eps)^{3/2}O_{z-2\eps}(1),
\end{eqnarray*}
where recall that since we are using coordinates in $U_j$, 
$f_{j+1}(z)=2\eps+z^2+b_{j+1}z^3+c_{j+1}z^4+\dots$.

These equations give, on $\p D_{slit}$,
\begin{equation}
dg_j=-\frac2\pi\Re\left(d\log((f_{j,\{4\}}^{-1})')+\sqrt\eps O_z(1)dz\right)
=dg_{j,\{4\}} + \sqrt\eps O_z(1)dz
\end{equation}
and similarly $dg_{j+1}=dg_{j+1,\{4\}} + \sqrt\eps O_z(1)dz$.

Having found $dg_j$ and $dg_{j+1}$, the integral (\ref{int2}) becomes
\begin{equation}\label{s}
\oint h_j(dg_{j+1}-dg_j)=\oint_{\partial D_{slit}}\left(h_{j,\{4\}}+
z^{3/2}O_z(1)\right)\left(
dg_{j+1,\{4\}}-dg_{j,\{4\}}+\sqrt\eps O(1)dz\right).
\end{equation}
Since $h_j$ is bounded, $\sqrt\eps O(1)\oint h_{j,\{4\}}dz=O(\eps^{3/2})$
and similarly on the path of integration the only singularity of
$dg_{j+1,\{4\}}$ or $dg_{j,\{4\}}$ occurs near the origin, where the
functions are at worst $O(z^{-1})$ so
$$\oint z^{3/2}O_z(1)\left(dg_{j+1,\{4\}}-dg_{j,\{4\}}\right)=\oint z^{1/2}O_z(1)dz = O(\eps^{3/2}).$$
Thus we can pull the errors in (\ref{s}) out of the integral,
giving $O(\eps^{3/2})$. Therefore the
integral (\ref{int2}) only depends on $b_j,c_j,b_{j+1},c_{j+1}$
up to an error $O(\eps^{3/2})$.

Lastly we estimate the third integral in (\ref{3ints}). 
As was the case for the first integral, on the two halves of $N_\delta(0)$
the integral of $dg_{j+1}$ tends to zero with $\delta$.
Now on $N_\delta(2\eps)$ we have $dg_{j+1}=O(\delta)$, and 
$h_{j+1}$ is constant ($1$ or $-1$)
on the two ``sides'' of $[0,2\eps]$; so 
\begin{eqnarray}\label{ss}
\oint_Xh_{j+1}dg_{j+1}&=&o(1)+\int_\delta^{2\eps-\delta}dg^+_{j+1}-
\int_{2\eps-\delta}^\delta dg^-_{j+1}\\
\nonumber
&=&o(1)+g_{j+1}(\delta)^+ -g_{j+1}(2\eps-\delta)^+-g_{j+1}(2\eps-\delta)^-
+g_{j+1}(\delta)^-
\end{eqnarray}
where the $+$ and $-$ refer the the two limits from above and below the axis,
and the $o(1)$ tends to zero with $\delta$.
Using $g_{j+1}(z)=\frac2\pi\Re\log (f_{j+1}^{-1})'(z)$, we see that the values
at $\delta$ and $2\eps-\delta$ of $g_{j+1}$ depend only on the $4$-jet
up to error $O(\eps^{3/2})$.

We have shown that the change in energy depends only on the $4$-jet of $f_j$
and $f_{j+1}$. 
Now when $j$ is not within $K$ of $0$ or $N$, $b_{j+1}=b_j+O(\eps)$ and 
$c_{j+1}=c_j+O(\eps)$ by Lemma \ref{smallj}. Changing $b_{j+1}$ and $c_{j+1}$
by $O(\eps)$ changes 
the second and third integrals (\ref{s}) and (\ref{ss}) by at most $O(\eps^{3/2})$;
so when $K<j<N-K$
the change in energy in fact depends only on the $4$-jet of $f_j$.

When $j<K$ or $N-K<j$, the energy depends on the $4$-jet of
{\it both} $f_j$ and $f_{j+1}$. However in these cases both $f_j$ and $f_{j+1}$
are independent of $U$ up to $O(1)$; 
in the next section we'll see that the energy depends
only on $j$ (through the Schwarzian derivative of $\sqrt{f_j}$) up to error 
$o(1/j)$.
This completes the proof that the change in Dirichlet energy only depends
on the $4$-jet of $f$ at the origin, up to terms in $o(\eps)$.

\subsection{Computation for a specific family of functions}
At this point we could simply work out the integrals
(\ref{s}) and (\ref{ss}), but it is less computationally painful
to do an explicit calculation for a particular family of functions.
This will also give a more accurate estimate of the energy when $j$
is near $0$ or $N$.

For $p,q\in i\R$ and $\{0,p,q\}$ distinct,  define
\begin{equation}\label{Fpq}
f_{p,q}(z)=2\sqrt{\frac qp}\int_0^zu\sqrt{\frac{u-p}{u-q}}du.
\end{equation}

\begin{figure}[htbp]
\begin{center}
\PSbox{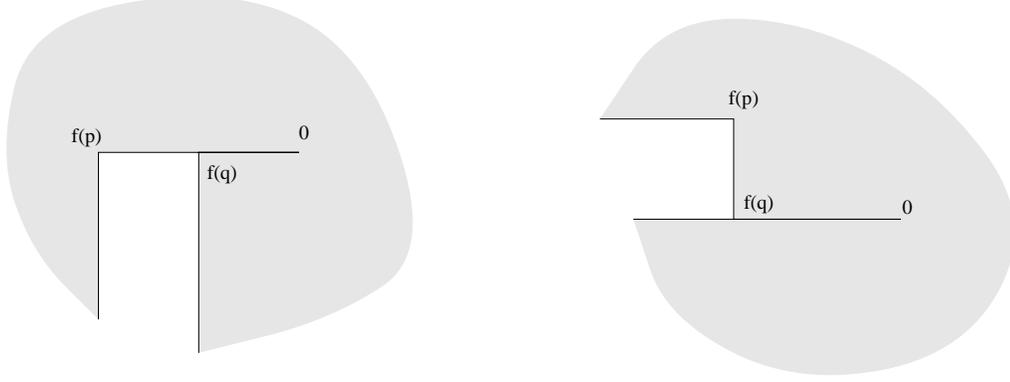}{-5in}{3in}
\end{center}
\caption{\label{pqelbow}The image of the RHP under the function $f_{p,q}$
is shown in gray.}
\end{figure}

If $-ip>0>-iq$ and $|p|>|q|$ this is a map from the RHP injectively
onto the region $U_{p,q}$
shown in gray in Figure \ref{pqelbow}a;
this follows from the Schwarz-Christoffel formula \cite{Ahl} ($f$ maps
the RHP to the polygonal region with angle $2\pi$ at the origin, $3\pi/2$ at
$f(p)$, $\pi/2$ at $f(q)$, and $2\pi$ at $\infty$.
If $-ip>0>-iq$ and $|p|<|q|$ this is still a well defined map which is 
not injective on $RHP$ (although it is locally injective).
This lack of injectivity is of no consequence to us. 
When $p$ and $q$ are on the same side of the origin the image
is as shown in Figure \ref{pqelbow}b. Again the map may or may not be injective
according to whether or not $p$ is between $q$ and $0$.

The $4$-jet of $f_{p,q}$ at the origin is
\begin{equation}
\label{expFpq}
f_{p,q}(z)=z^2 + \left(\frac1q-\frac1p\right)\frac{z^3}3+
\left(\frac3{4q^2}-\frac1{2pq}-\frac1{4p^2}\right)\frac{z^4}4+O(z^5).
\end{equation}
For any $f$ of the form (\ref{expF}) there is
a $p,q$ for which $f_{p,q}$ has the same $4$-jet at the origin as $f$,
provided that $b\neq 0$:
it suffices to take $p=\frac{12b}{16c-27b^2}$ and
$q=\frac{12b}{16c+9b^2}$.
Note that when $p,q$ take these values, 
two of the points $\{0,p,q\}$ are equal only when $b=0$
(the cases when one of $p,q$ is infinite are allowed).
The case $b=0$ will be dealt with later.

The integral in (\ref{Fpq}) can be explicitly evaluated, giving
$$f_{p,q}(z)=\frac14\sqrt{\frac qp}\left(\sqrt{(z-p)(z-q)}(4z-2p+6q)-
2(p^2+2pq-3q^2)\log\left(\sqrt{z-p}+\sqrt{z-q}\right)\right)-C_2$$
where $C_2$
is chosen so that $f_{p,q}(0)=0$:
$$C_2= \frac14\sqrt{\frac qp}\Bigl(\sqrt{pq}(-2p+6q)-
(p^2+2pq-3q^2)\log\left(-(\sqrt{p}-\sqrt{q})^2\right)\Bigr).$$
We have explicitly
$$f(p)= -\frac14q(2p-6q)+\frac14\sqrt{\frac qp}\Bigl(p^2+2pq-3q^2\Bigr)\log\left(
\frac{\sqrt p+\sqrt q}{\sqrt q-\sqrt p}\right)$$
and
$$f(q)= -\frac14q(2p-6q)+\frac14\sqrt{\frac qp}\Bigl(p^2+2pq-3q^2\Bigr)\log\left(
\frac{\sqrt p+\sqrt q}{\sqrt p-\sqrt q}\right).$$
Thus
\begin{equation}\label{p-q}
f(p)-f(q)=\frac{\pi i}4\sqrt{\frac qp}(p^2+2pq-3q^2)
\end{equation}
where we chose the sign of the square root and branch of log to correspond
to the situation of Figure \ref{pqelbow}.

Suppose $-ip>0>-iq$.
Then the $\delta$-normalized Dirichlet energy for the region $U_{p,q}=
f_{p,q}(RHP)$ is
$\oint_{U'} h dg $ where $h$ is $0$ on the vertical boundary from
$-i\infty$ to $f(q)$, 
$-1$ on the horizontal boundary from $f(q)$ to $0$,
$1$ on the boundary
from $0$ to $f(p)$, and $2$ on the vertical boundary from $f(p)$
to $\infty$ (note that $h$ has different values on the two ``sides'' of $[f(q),0]$). 

To compute $E_\delta$, we pull $h$ back to RHP. The preimage of $N_\delta(f(p))$
is to first order a disk around $p$ (that is, it converges to a round disk
when $\delta\to0$); let $\delta_p$ denote its radius.
Similarly let $\delta_q$ and $\delta_0$ be the radii of the preimages of the
disks around $f(q)$ and $f(0)$, respectively. Let $\delta_\infty$
be defined by: $1/\delta_\infty$ is the radius of the preimage of the
disk of radius $1/\delta$ around $0$. 

Then $E_\delta$ is (recalling that $g$ is constant up to lower order terms
on the boundary of the $\delta$-neighborhood of the singularities)
\begin{eqnarray}E_\delta&=&
-\int_{q+i\delta_q}^{-i\delta_0} d(g\circ f_{p,q}) + \int_{\delta_0}^{p-i\delta_p}
d(g\circ f_{p,q})+2\int_{p+i\delta_p}^{1/\delta_\infty} d(g\circ f_{p,q})\nonumber\\
\label{2ggg}
&=&-2g(f_{p,q}(\delta_0))+g(f_{p,q}(q+i\delta_q))-g(f_{p,q}(p+i\delta_p))
+2g(f_{p,q}(1/\delta_\infty)).
\end{eqnarray}

From (\ref{expFpq}) we have $\delta_0^2=\delta$ up to higher order terms.
For $z$ close to $p$ we have from (\ref{Fpq})
$$f_{p,q}(z)=f_{p,q}(p) + \frac43\frac{\sqrt{qp}}{\sqrt{p-q}}(z-p)^{3/2}+
O((z-p)^2)$$
and so up to higher order terms we have
\begin{equation}\label{delp}
\delta_p=\left(\frac34\delta\sqrt{\frac{p-q}{pq}}\right)^{2/3}
\end{equation}
and similarly  near $q$
$$f_{p,q}(z)=f_{p,q}(q) + 4\sqrt{\frac{q}p}q\sqrt{q-p}(z-q)^{1/2}+
O(z-q)$$ so
\begin{equation}\label{delq}
\delta_q=\left(\frac\delta4\sqrt{\frac pq}\frac1{q\sqrt{q-p}}\right)^2.
\end{equation}
For large $z$ we have $f'_{p,q}(z)=2\sqrt{\frac{q}p}z+O(1)$
which implies $f_{p,q}(z)=\sqrt{\frac{q}p}z^2+O(z)$
and so
\begin{equation}\label{delinf}
\delta_\infty=\left(\delta\sqrt{\frac qp}\right)^{1/2}.
\end{equation}

Plugging these in to (\ref{2ggg}) with $g\circ f_{p,q}=-\frac2\pi
\Re\log f_{p,q}'$
gives the Dirichlet energy to be
\begin{eqnarray}\nonumber
E_\delta&=&-2\left(-\frac2\pi\Re\log\left(\delta^{1/2}\sqrt{\frac pq}\right)\right)
-\frac2\pi\Re\log\left(\frac{q\sqrt{q-p}}{\frac\delta4\sqrt{\frac pq}
\frac1{q\sqrt{q-p}}}
\right)\hskip1in\\
&&\hskip1in+\frac2\pi\Re\log\left(\frac{p}{\sqrt{p-q}}\left(\frac34\delta\sqrt{
\frac{p-q}{pq}}\right)^{\frac13}\right)-\frac4\pi\Re\log\left(
\frac1{\delta^{1/2}}\left(\frac qp\right)^{1/4}\right)\\
&=&\label{DEpq}
\frac{20}{3\pi}\log\delta+\frac1{3\pi}\log\left(\frac{9p^{11}}{4^8q^{19}|p-q|^8}
\right).
\end{eqnarray}

By symmetry we get the same energy when $-ip<0<-iq$.
A similar calculation shows that 
the same energy is obtained in the remaining cases when $p$ and $q$
are on the same side of the origin. 

When we extend the cut by a small amount
$2\eps$, the new region has a new uniformizing
function of the same form $f_{p',q'}$, where 
$p'$ and $q'$ are defined by changing $p$ and $q$ 
so that $f(q)$ and $f(q)$ each change by $-2\eps$.
Let $dp$ and $dq$ denote the changes in $p$ and $q$ for an infinitesimal
$\eps$.
We must first have $0=d(f(p)-f(q))$.
This relates the change in $p$ to the change in $q$. 
From (\ref{p-q}) we have the equation
\begin{eqnarray*}0&=&d(f(p)-f(q))\\
&=&\left(\sqrt{\frac qp}(2p+2q)+(p^2+2pq-3q^2)
(\frac{-\sqrt q}{2p^{3/2}})\right)dp+
\left(\sqrt{\frac qp}(2p-6q)+\frac 1{\sqrt{pq}}(p^2+2pq-3q^2)\right)dq,
\end{eqnarray*}
or
$$dq = -\frac qp\left(\frac{3p^2+2pq+3q^2}{p^2+6pq-15q^2}\right)dp.$$
Since $\sqrt{\frac qp}(p^2+2pq+3q^2)$ does not change, we have $d(f(p))=$
$$-\frac q2dp+(3q-\frac12p)\, dq+\frac14\sqrt{\frac qp}(p^2+2pq-3q^2)
\left(\left(
\frac1{\sqrt p+\sqrt q}-\frac1{\sqrt p-\sqrt q}\right)\frac{dp}{2\sqrt p}
+\left( \frac1{\sqrt p+\sqrt q}+\frac1{\sqrt p-\sqrt q}\right)
\frac{dq}{2\sqrt q}\right)$$
$$=\frac{-16pq^2}{(p^2+6pq-15q^2)}dp.$$

Now the change in $E_\delta$ when $d(f(p))=-2\eps$ and
$d(f(p))=d(f(q))$ is 
$$-2\eps\frac{dE_\delta}{dp}\frac{dp}{d(f(p))}=\frac{4(5p+7q)(p-q)}{\pi p(p^2+6pq-15q^2)}
\cdot\frac{(p^2+6pq-15q^2)}{-16pq^2}
$$
$$=\frac{\eps(5p+7q)(p-q)}{2\pi p^2q^2}.$$

Finally the Schwarzian derivative of $\sqrt{f_{p,q}}$
is 
$$\frac34\left(\frac3{4q^2}-\frac1{2pq}-\frac1{4p^2}\right)-
\frac14(\frac1q-\frac1p)^2=\frac{(5p+7q)(p-q)}{16p^2q^2}$$
which is $\frac\pi8$ times the change in $E_\delta$.
This completes the proof when $j\in[K,N-K]$ except in the case $b=0$. 
For the case $b=0$ see the function $f_q$ of section \ref{begedge} below.
The case $j\not\in[K,N-K]$ is dealt with in the next section.
\hfill{$\square$}

\subsection{Beginning and ending of a cut}
As above, the change in energy near the beginning of a cut only depends
on the germ of $f$ near the tip of the cut. 
Near the beginning of the cut, the germ of $f$ only depends on
whether or not the cut starts at a corner or at an edge (Lemma \ref{smallj}).
We computed the limiting functions $f$ when $j$ is near the beginning
or ending of the cut in the proof of Lemma \ref{nextdom}.

\subsubsection{beginning on an edge}\label{begedge}
Suppose first that the cut begins on a straight edge of $U$.
We compute the change in $E_\delta$ between the time when there is no cut to the time when
the cut has length $2\eps j$.

Let $h_q$ be the function on $RHP-[0,q]$ whose boundary values
are $0$ on the $y$ axis, $1$ on the upper boundary of the cut $[0,q]$
and $-1$ on the lower boundary of the cut. 
Up to an additive constant, this is the height function on $U$
near the beginning of the cut.

We can compute the $\delta$-normalized Dirichlet energy of $h_q$ in
the same way as we did the functions $f_{p,q}$ of the previous
section.  Specifically, the map from RHP to $U_q=RHP-[0,q]$ is
\begin{eqnarray*}
f_q(z)&=&\sqrt{2q z^2+q^2}-q\\
&=&z^2-\frac1{2q}z^4+O_z(z^6),
\end{eqnarray*}
and the Dirichlet energy is
\begin{equation}\label{Fqener}
2g\circ f_q(\delta_0)-g\circ f_q(\sqrt{-\frac{q}2}+\delta_1)-
g\circ f_q(-\sqrt{-\frac{q}2}+\delta_2),
\end{equation}
where (to first order) $f_q$ maps the $\delta_0$-neighborhood of the origin to 
the $\delta$-neighborhood of the origin, and the 
neighborhoods of $\pm \sqrt{-\frac q2}$ of radius $\delta_1$ to
the $\delta$-neighborhood of $-q$.

As before $\delta_0=\delta^{1/2}$; one computes 
$\delta_1=\delta_2=\frac{\delta^2}{4q\sqrt{q/2}}$. Plugging into (\ref{Fqener})
with $g\circ f_q=-\frac2\pi\Re\log f_q'$
gives
\begin{equation}\label{6pi}
E_\delta=\frac6\pi\log\frac1\delta+\frac6\pi\log q+\frac2\pi\log 2.
\end{equation}

When $q=2\eps j$, i.e. $U_q=RHP-[0,2\eps j]$, this gives the energy
change due to the first $j$ steps of the cut.
The change in Dirichlet energy between
$j$ and $j+1$ is $\frac6{\pi j}$ which is 
$-\frac{8\eps}\pi S\sqrt{f_j}(0)$ as required.

Note that if we set $\delta=\eps$ in (\ref{6pi}), the $\eps$-normalized
energy due to the beginning of the cut is
$$\frac6\pi\log j + const+ O(\eps).$$
In particular when $j=K=\xi/\eps$ this is $\frac{6}\pi\log\frac1\eps+O(1)$.

\subsubsection{starting at a concave corner}
In case the cut starts at a corner of $U$ (of angle $3\pi/2$),
the change in energy can be computed in a similar manner.
Again we compute the change in energy between the time when
there is no cut to the time when the cut has length $2\eps j$.

Let $TQP$ be the three-quarters plane, $TQP=\{(x,y)|x>0\mbox{ or }y>0\}$ and 
define $U_q=TQP-[0,q]$.  We first compute $E_\delta$ on $TQP$.
We can take 
the average height function $h$ on $TQP$ to be $1$ on the negative
$x$-axis and $0$ on the negative $y$-axis, that is,
$h=const+\frac2{3\pi}\Im\log(z)$. Then
$E_\delta(h)$ on $TQP$ is $\oint h\,dg = g(\delta)-g(\frac1\delta)$ and
up to higher order terms $g(\delta)=\frac2{3\pi}\log\frac1\delta$
and $g(\frac1\delta)=-\frac2{3\pi}\log\frac1\delta$.
So the energy is $\frac4{3\pi}\log\frac1\delta$.

The map from $RHP$ to $U_q$ is
$$f_q(z)=2\sqrt{-q}\int_0^z\frac{wdw}{\sqrt{w-q}}=2\sqrt{-q}
\left(\frac23(w-q)^{3/2}+
2q(w-q)^{1/2}\right)+\frac{8q^2}3.$$
At the origin we have
$$f_q(z)=z^2+\frac1{3q}z^3+\frac{3}{16q^2}z^4+O(z^5).$$

Now 
$E_\delta=\oint_{U_q'} h\,dg=-g\circ f_q(\frac1{\delta_\infty})
-g\circ f_q(q+i\delta_q)+2g\circ f_q(\delta_0)$.
Again $\delta_0=\delta^{1/2}$ and 
near $\infty$, 
$$\delta^{-1}=f(\frac1{\delta_\infty})=\frac{4\sqrt{-q}}3\delta_\infty^{-3/2}.$$
Near $q$ we have
$f(z)=f(q)+4q\sqrt{-q}\sqrt{z-q}+O(z-q)$, so
$$|4q\sqrt{-q}|\delta_q^{1/2}=\delta.$$
Plugging all this in we have 
$$E_\delta(h_q)= 
\frac2\pi\log\left(2\sqrt q\left(\frac3{4\sqrt q\delta}\right)^{1/3}\right)+
\frac2\pi\log\left(\frac{8q^3}{\delta}\right)-
2\frac2\pi\log\left(2\sqrt{\delta}\right)$$
$$=\frac{14}{3\pi}\log\frac1\delta+\frac{20}{3\pi}\log q+const.$$
The difference in energy between $U_q$ and $TQP$ is then
$$\frac{10}{3\pi}\log\frac1\delta+\frac{20}{3\pi}\log q+const.$$
Now $f(q)=\frac{8q^2}3,$ which is $-2\eps j$ when
$q=\sqrt{\frac{3\eps j}4}.$
Plugging this value of $q$ in gives the energy due to the cut of $P_j$ to be
\begin{equation}\label{concavecorner}
\frac{10}{3\pi}\log\frac1\delta+\frac{10}{3\pi}\log (\eps j)+const.\end{equation}

When $j$ changes by $1$ the energy changes by $\frac{10}{3\pi j}$,
which is again $-\frac{8\eps}\pi S\sqrt{f_q}(0)$.

Setting $\delta=\eps$ in (\ref{concavecorner}) gives
the contribution to $E_\eps$ from the beginning of the cut as
$\frac{10}{3\pi}\log j + const + O(\eps)$.

\subsection{Ending on the interior of an edge}
A similar computation holds in this case, with $f_q$ as in the proof of Lemma
\ref{nextdom}. However there is a shorter method, obtained as follows.

The change in $E_\delta$ near the end of a cut
only depends on the local structure of the average height function
near the end of the cut. Therefore the energy can be obtained
from the known value of the energy for a rectangle.
Indeed, cutting a rectangle into two rectangles with a single edge,
the change in $E_\eps$ due to the beginning of the cut,
plus the change due to the ending of the cut, plus the contribution
from the ``central" terms, must equal total energy change which is 
$\frac{24}{\pi}\log\frac1\eps+const+O(\eps)$
(see (\ref{rectenergy})). The contribution at the beginning
of the cut is $\frac6\pi\log\frac1\eps+const+O(\eps)$ (see section \ref{begedge}), 
so the contribution at the end of the cut
is $$\frac{18}\pi\log\frac1\eps+const+O(\eps)=-\frac{48}\pi\left(
-\frac38\log\frac1\eps+const+O(\eps)\right).$$ The constant is independent
of $U$ by Lemma \ref{smallj}.

\subsection{Ending at a concave corner}
The energy for the end of a cut ending at a corner of $U$ can
be computed from the previous case. Given an ell-shaped region, 
cut it into two rectangles with a single cut beginning
on an edge and ending at the corner. 
We can compute the change in Dirichlet energy by starting the cut
at the concave corner and ending at the edge; the contribution
from the start of the cut starting at the concave corner is
(cf. (\ref{concavecorner})) $\frac{10}{3\pi}\log\frac1\eps+const+O(\eps)$
and the contribution from the end of this cut is
$\frac{18}\pi\log\frac1\eps+const+O(\eps)$ by the previous paragraph.
The sum of these must equal the contribution for starting from
the edge ($\frac6\pi\log\frac1\eps+const+O(\eps)$) and ending at the concave corner.
Therefore the contribution due to a cut ending at a concave
corner is 
$$\frac{46}{3\pi}\log\frac1\eps+const+O(\eps)=
-\frac{48}\pi\left(-\frac{23}{72}\log\frac1\eps+const+O(\eps)\right).$$

This completes the proof of Lemma \ref{changeinDE}.

\section{The case of rectangles}\label{rectsect}
Let $P$ be a $(2m-1)\times (2n-1)$ rectangle with  a unit square
removed at the lower left corner.
By Temperley's trick \cite{Temp}, domino tilings 
of $P$ are in bijection with spanning trees of the $m\times n$ 
grid graph, rooted at the lower left corner.

The number of spanning trees of the $m\times n$ grid, which by 
Kirchhoff \cite{Kirch}
is the product of the non-zero eigenvalues of the Laplacian, is
$$\prod_{(j,k)\neq0}\left(4-2\cos\left(\frac{\pi k}n\right)-
2\cos\left(\frac{\pi j}{m}\right)\right),$$
where $j$ (resp. $k$) runs from $0$ to $m-1$, resp. $0$ to $n-1$.
The log of this formula was asymptotically computed in \cite{Dup}:
\begin{prop}[\cite{Dup}]\label{rectform}
The log of the number of spanning trees of an $n\times m$ rectangle is
$$\frac{4Gmn}\pi+(m+n)\log(\sqrt2-1)-\frac12\log(m)
+\log(\eta(e^{-2\pi n/m}))-\frac14\log(2)+O(\frac1{mn}).$$
\end{prop}

Here $G$ is Catalan's constant 
and $\eta$ is the Dedekind eta-function
$$\eta(q)=q^{1/24}\prod_{k=1}^\infty(1-q^k).$$

The area of $P$ is $(2n-1)(2m-1)-1=4nm-2n-2m$, and the perimeter
is $4n+4m-4$, so this formula may be rewritten
\begin{equation}\label{realform}
\frac{G}{\pi}\mbox{Area}(P) +
\left(\frac{G}{2\pi}+\frac{\log(\sqrt{2}-1)}4\right)
\mbox{Perim}(P)
-\frac12\log(m)+\log(\eta(e^{-2\pi n/m}))+C+ O(m^{-2}),
\end{equation}
where 
$$C= \log\left(\frac{2^{5/4}}{1+\sqrt2}\right)+ \frac{2G}\pi.$$

\subsection{Dirichlet energy for a rectangle}
Let $U$ be the rectangle $[0,\frac12]\times[0,\frac\tau2]$ in $\C$,
with base point at the lower left corner.
The average height function $h$ for $U$ is, up to an additive constant,
the harmonic function which is $0,1,2,3$ on the lower, right, upper and 
left boundaries respectively. 

The function $h$ has an explicit expression in terms of the 
the Weierstrass
elliptic function $\wp(z)=\wp_{1,i\tau}(z)$ (see \cite{Ahl}: this is the doubly-periodic
function with periods $1$ and $i\tau$ and a double pole at each point of the lattice).
\begin{lemma} Up to an additive constant we have
$$h(z) = -\frac2\pi\Im\log\wp'(z).$$
\end{lemma}

\begin{proof} From \cite{Ahl} we have
$$\wp'(z) = \sum_{w\in\Gamma}\frac{-2}{(z-w)^3},$$
where $\Gamma=\Z + i\tau\Z.$
Note that $\wp'(z)$ is real when $z\in[0,\frac12]$ or $z\in[0,\frac12]+i
\frac{\tau}2$ and
pure imaginary when $z\in i[0,\frac{\tau}2]$ or $z\in i[0,\frac{\tau}2]+
\frac12$.
Furthermore on a fundamental domain for $\Gamma$, 
$\wp'(z)$ is zero or infinite only at the corners of $U$ (see \cite{Ahl}).
So the argument of $\wp'(z)$ is constant on each edge of the rectangle $U$.
At a corner $z_j,~z_j\neq 0$, we have $\wp'(z)=c_j(z-z_j) + O((z-z_j)^2)$ 
and so 
$\frac{-2}\pi\Im\log\wp'(z)$ changes by $+1$ at each corner of $U$ when
going counterclockwise around $U$.
\end{proof}

Let $U'$ be $U$ minus the $\delta$-neighborhood of
the four corners of the rectangle.
Then 
$$E_\delta(h)=\int_{\partial U'}h\,dg=\int_\delta^{1/2-\delta}0\,dg+
\int_{1/2+i\delta}^{(1+i\tau)/2-i\delta}1\,dg+
\int_{(1+i\tau)/2-\delta}^{i\tau/2+\delta}
2\,dg+\int_{i\tau/2-i\delta}^{i\delta}3\,dg+O(\delta)$$
where the contribution on the boundaries of the neighborhoods
of the corners is $O(\delta)$.
Since when $\delta$ is small $g$ is essentially constant
on the neighborhoods of radius $\delta$ of the corners, this energy is
$$E_\delta(h)=-g(\frac12-\delta)-g(\frac12+i\frac{\tau}2-\delta)-g(i\frac{\tau}2
+\delta)+3g(\delta)+O(\delta).$$

At the origin, $\wp'(z)=-\frac{2}{z^3}+O(1)$ and so 
$$\frac2\pi\log\wp'(\delta e^{i\theta}) = 
\frac2\pi\log\left(\frac{-2}{\delta^3e^{3i\theta}}+const+O(\delta)\right)$$
$$= \frac{6}\pi\log\frac1\delta -\frac{6\theta i}\pi+\frac2\pi\log(-2)+ O(\delta^3).$$ 
Near the other three corners of $U$, 
$$\frac2\pi\log\wp'(\delta e^{i\theta}-z_j)=
-\frac{2}\pi\log(c_j\delta e^{i\theta})+O(\delta)$$
where $c_j=\wp''(z_j)$.
Using $g(z)=\frac2\pi\Re\log\wp'(z)$,  the energy is
$$E_\delta(h)=-\frac2\pi\left(\log(c_1\delta)+\log(c_2\delta)+\log(c_3\delta)\right)
+\frac{18}\pi\log(\frac1\delta)$$
$$=\frac{24}\pi\log\frac1\delta -\frac2\pi\log(c_1c_2c_3),$$
where $c_1,c_2,c_3$ are the derivatives $\wp''(z_j)$ at the three
other corners of $U$.

\begin{lemma}
We have 
$$c_1c_2c_3=\frac12(2\pi)^{12}\eta(e^{-2\pi\tau})^{24}.$$
\end{lemma}

\begin{proof}
From the differential equation
$$(\wp'(z))^2=4(\wp(z)-e_1)(\wp(z)-e_2)(\wp(z)-e_3)$$
(where $e_j=\wp(z_j)$)
we obtain (upon differentiating the logarithms of both sides)
$$2\frac{\wp''(z)}{\wp'(z)}=\frac{\wp'(z)}{\wp(z)-e_1}+\frac{\wp'(z)}{\wp(z)-e_2}+
\frac{\wp'(z)}{\wp(z)-e_3}$$
from which we get
$$\wp''(z_1)=2(e_1-e_2)(e_1-e_3),$$
with similar expressions for $c_2,c_3$. Their product is
$$c_1c_2c_3=8(e_1-e_2)^2(e_1-e_3)^2(e_2-e_3)^2$$
and by \cite{Apostol} this equals
$$\frac12\Delta = \frac12(2\pi)^{12}\eta(e^{-2\pi\tau})^{24}$$
\end{proof}

The energy is therefore
$$\frac{24}\pi\log\frac{1}\delta-\frac2\pi\log\left(
\frac12(2\pi)^{12}\eta(e^{-2\pi\tau})^{24}\right).$$
This is the $\delta$-energy for a $\frac12\times\frac\tau2$ rectangle.

Let $U_{\alpha,\beta}$ be the rectangle $[0,\alpha]\times[0,\beta]$,
where $\tau=\beta/\alpha$. We can scale $U_{\alpha,\beta}$ by $\frac1{2\alpha}$ 
to get $U$.
As a consequence the above expression is the $2\alpha\delta$-energy for 
$U_{\alpha,\beta}$. To get the $\delta$-energy for $U_{\alpha,\beta}$, 
substitute $\delta/(2\alpha)$ for $\delta$ in the above expression.
We get
\begin{equation}\label{rectenergy}
E_\delta(U_{\alpha,\beta})=\frac{24}\pi\log\frac{2\alpha}\delta-\frac2\pi\log\left(
\frac12(2\pi)^{12}\eta(e^{-2\pi\tau})^{24}\right).
\end{equation}

\begin{prop}\label{epsrectform} Let $U$ be a rectangle;
let $P_\eps$ be a sequence of Temperleyan rectangles of area $A_\eps$
and perimeter $\mbox{Perim}_\eps$ approximating $U$.
Then the log of the number of domino tilings of $P_\eps$ is
$$\frac{GA_\eps}{\pi\eps^2} +
\left(\frac{G}{2\pi}+\frac{\log(\sqrt{2}-1)}4\right)
\frac{\mbox{Perim}_\eps}{\eps}
-\frac\pi{48}r_2(\eps,U) + C + O(\eps^2),$$
where $r_2(\eps,U)$ 
is the $\eps$-normalized Dirichlet energy of the average height function
(given by putting $\eps$ for $\delta$ in (\ref{rectenergy}))
and $C$ is a universal constant.
\end{prop}

\begin{proof}
If $U$ is an $\alpha\times\beta$ rectangle, this follows from 
(\ref{realform}) upon setting $2m-1=\alpha/\eps,~2n-1=\beta/\eps$.
\end{proof}

This proposition tells us why the `natural' choice of $\delta$
in $E_\delta(h)$ is $\delta=\eps$ or some constant multiple of $\eps$. 
In fact there is a multiple which makes the constant $C$ in the above
formula vanish, although we will not need this accuracy. 

\section{Loop-erased random walk}\label{LERW}
The loop-erased random walk in a finite graph $G$ has the following simple
description. Let $b_0,b_1$ be two vertices of $G$. Take a simple
random walk starting from $b_0$ and stopping at $b_1$. Erase from the path
its loops, in chronological order. That is, if there is a loop,
erase the first loop 
(the first time the path comes back to the same vertex twice).
For the new path, if there is still a loop, erase the first loop, and so on.
The remaining path is a simple path from $b_0$ to $b_1$.

There is a well-known connection between the LERW and spanning trees 
\cite{Pem}: in a uniformly chosen spanning tree on a region
$P\subset\Z^2$, the unique arc (branch) from $a$ to $b$
has the same distribution as the LERW from $a$ to $b$ on the same region. 
Pemantle also showed that on an infinite domain such as $\Z^2$ or the upper
half of $\Z^2$, the LERW is well-defined (despite recurrence of the
simple random walk. See \cite{Pem}).

\subsection{LERW and dominos}
Let $P$ be a $2n-1\times 2n-1$ square Temperleyan polyomino with base square 
$b_0$ on the center of the right edge. 
Recall how Temperley's trick works:
a domino tiling of $P$ corresponds to a spanning tree of the graph $H(P)$
which is rooted at $b_0$ in the following way. Each 
domino covering a vertex $v$ of $H(P)$ has
white vertex covering the center of an adjacent edge $e$ of $H(P)$. 
In the associated spanning tree the outgoing edge from $v$
points in the direction $e$. The union of these edges forms
a spanning tree with all edges oriented towards the root at $b_0$.
 
Let $Q$ be the region obtained from $P$ by removing
a single black square $b\in B_0$ and a single white square $w$.

\begin{lemma}\label{2holecomb} When $b$ is on the boundary of $P$,
domino tilings of $Q$ are in bijection 
with spanning trees of $P$ for which the branch from
$b$ to $b_0$ contains the edge $w$ (traversed in either direction).
\end{lemma}

\begin{proof} 
Let $T$ be a domino tiling of $Q$.
Let $b'$ and $b''$ be the 
neighboring squares of $w$ which are in $H$.
Temperley's trick assigns to a vertex $x\in H$ an outgoing edge
$e$ if and only if $xe$ is a domino of $T$.
Temperley's trick, applied to $T$, gives a set of directed
edges of the graph $H$ such that each vertex 
has exactly one outgoing edge, except for $b_0$ and $b$
which have no outgoing edges. Furthermore the union of these edges is a forest,
i.e. each component is a tree. 
There are exactly two components to this graph
since each component is rooted at exactly one of $b_0$ or $b$.

We claim that $b'$ and $b''$ are in different components of this forest.
To see this, first note that the directed branch from $b'$ 
cannot pass through $b''$,
for otherwise the path from $b'$ to $b''$ followed by the segment
from $b''$ to $w$ to $b'$ would be a lattice path in $H$ enclosing 
an odd number of squares of $Q$
\footnote{From the Euler formula for a disk $F-E+V=1$, for any simple
closed path in $\Z^2$ the sum of the number of faces, edges and vertices
strictly enclosed is odd.}
and so could not arise from a domino tiling of $Q$. Similarly the paths from
$b'$ and $b''$ cannot end both at $b_0$ or both at $b$
because their union would contain a closed lattice path from $b'$ to $b''$.

So the paths from $b'$ and $b''$ end one at $b_0$ and one at $b$.
On the path which ends at $b$, shift the dominos along it by
one square towards $b$, and add an extra domino from
$w$ to its freed neighbor.
This makes a domino tiling of $P$ in which the tree branch from 
$b$ passes through $w$ (shifting dominos by $1$ along a tree branch has the effect
of changing the direction of the edges on that branch).

This process is reversible: from a tiling of $P$ whose branch from
$b$ to $b_0$ passes through $w$, shift the dominos by one up to $w$
to get a tiling of $Q$.
\end{proof}

When $b$ is not on the boundary of $P$, it is possible that 
one of the paths from $b'$ or $b''$ winds around $b$, or that 
these two paths both lead to $b_0$, enclosing $b$. So the lemma
does not hold in that case. 

By Lemma \ref{2holecomb}, 
to prove Theorem \ref{lerwthm}
it suffices to be able to count the number of tilings of $Q$,
or rather, to compute the ratio of the number of tilings of $Q$
to the number of tilings of $P$.

\subsection{Region with a white hole}\label{whitehole}
First we compute the coupling function on $Q$.
Since $Q$ has a hole (the hole $w$) which does not enclose the same
number of black and white squares modulo $2$, 
the weighted adjacency matrix $A$ of $Q'$ (the dual graph of $Q$) is not
a Kasteleyn matrix for $Q$.
In \cite{Kast2}, Kasteleyn describes how to redefine the weights of $A$
to get a Kasteleyn matrix in this case (rather, he describes how to
get weights for general
planar graphs).  One way to get a Kasteleyn matrix for $Q'$ is the following. 
Start with the weighted adjacency matrix $A=A(Q')$ of section \ref{confprops}.
Take a path $\gamma$ of vertices in $Q$ from the outer boundary
to a vertex of $w$ ($\gamma$ is a path of {\em faces} of $Q'$).
Every edge in $\gamma$ crosses an edge $v_1v_2$ of $Q'$.
For each edge in $\gamma$, change the sign on the corresponding 
matrix entries $A_{v_1v_2}$ and $A_{v_2v_1}$. 
The matrix $A'$ with these new signs is a Kasteleyn matrix for $Q'$.
In particular it has the property that its determinant is the square
of the number of matchings, and its inverse is the coupling function 
$C_Q$ for $Q$ \cite{Kast2,Kenyon.prob}. 

Let $\tilde Q'$ be the double cover of the graph $Q'$, 
branched around the face containing $w$. That is, $\tilde Q'$
is defined by the property that a closed path in $Q'$ lifts to a closed
path in $\tilde Q'$ if and only if it winds an even number of
times around the face $w$. Similarly we define the double
cover $\tilde H$ of $H$ as follows. Recall that $w$ is an edge of $H$.
Remove $w$ from $H$ and let $\tilde H$ be the double cover of $H-\{w\}$
branched over the face which contained the edge $w$.
The double cover of $H'$, the dual of $H$, is similarly defined
(and denoted $\tilde H'$).

For each fixed $v$, the coupling function $C_Q(v,z)$ lifts to
a discrete analytic function $\tilde C_Q$ on $\tilde Q'$:
let $z'$ be a lift of $z$ and $v'$ a lift of $v$; then
define $\tilde C_Q(v',z')$ to be 
$\pm C_Q(v,z)$, with the $+$ sign when $v',z'$ are on the same sheet
of the cover (in the sense that there is a path from $v'$ to $z'$
which does not cross a lift of $\gamma$) and $-$ sign otherwise.

For a point $x\in Q'$ let $x'$ and $x''$ denote its two lifts.

Let $v\in W_1$ and $z\in B_1$. For lifts $v'$ of $v$ and $z'$ of $z$,
the real part of $\tilde C_Q(v',z')$ is harmonic on $\tilde H'$ except 
when $z=v\pm\eps$ or $z=w\pm\eps$ (as in (\ref{couplgreen1}), with extra
singularities at $w\pm\eps$).
So letting $\tilde G$ denote the Green's function on $\tilde H'$,
and $v_\pm = v \pm\eps,$ $w_\pm=w \pm\eps$, it must be that 
$\Re\tilde C_Q(v',z')$ is a linear combination of the Green's
functions $\tilde G(v_\pm',z'),\tilde G(v_\pm'',z'),\tilde G(w_\pm',z'),
\tilde G(w_\pm'',z')$. By the antisymmetry under changing sheets
and (\ref{couplgreen1}),
$\Re\tilde C_Q(v',z')$ is of the form
$$\tilde G(v_+',z')-\tilde G(v_-',z')-\tilde G(v_+'',z')+\tilde G(v_-'',z')+
\alpha_1\left(\tilde G(w_+',z')- \tilde G(w_-',z')\right)
+\alpha_2\left(\tilde G(w_+'',z')- \tilde G(w_-'',z')\right),$$
for some (real) constants $\alpha_1,\alpha_2$ which depend on $v$.
These constants are determined by the following condition on $C_Q$:
the harmonic conjugate of $\Re C_Q$ must be zero at the four points
$b_0',b_1',b_0'',b_1''$.
These four conditions give only $2$ linear equations for $\alpha_1,\alpha_2$ 
since the value at $b''_0$ is by construction
the negative of the value at $b'_0$ and similarly
for $b_1''$.

We need to show that the coefficients 
$\alpha_1,\alpha_2$ converge as $\eps\to 0$, and compute their limit.
To show this requires two steps. First,
we will show that the Green's functions $\tilde G(y,z)$
converge to the corresponding continuous Green's functions,
and moreover the terms $\frac1\eps(\tilde G(w_1',z)- \tilde G(w_1'',z))$ and
$\frac1\eps(\tilde G(w_2',z)-\tilde G(w_2'',z))$ converge to the derivatives of the
continuous Green's functions. These derivatives have simple
poles at the origin. Then we will show that there is a {\em unique}
pair of analytic functions $F_0,F_1$ with the required properties
having simple poles at the origin.

The convergence of the Green's functions on $\tilde H'$ is standard:
As $\eps\to0$, if $y'$ does not converge to $0$ then 
$\tilde G(y',z')$ has the form
$\tilde G(y',z')=\frac2\pi\log|y'-z'|+\frac2\pi\log\frac1\eps+const+ O(\eps/|z'-y'|)$
(see e.g. \cite{Spitzer}).
This is enough to conclude that as $\eps\to0$, 
$\frac1{\eps}(\tilde G(y-\eps,z)-\tilde G(y+\eps,z))$ tends to 
a harmonic function with zero boundary values and a singularity
at $y$ of the form $\Re\frac1{\pi(z-y)}$. This is the derivative 
$\frac{\partial}{\partial y}\tilde g(y,z)$ of the continuous Green's function 
$\tilde g(y,z)$.

From the symmetry $\tilde G(y,z)=\tilde G(z,y)$ we can conclude
the same when $y$ tends to $0$ as $\eps\to0$.
In conclusion $\frac1\eps(G(w_1',z)- G(w_1'',z))$ and 
$\frac1\eps(G(w_2',z)- G(w_2'',z))$ converge to harmonic functions with
simple poles at the origin (and single-valued harmonic conjugates). 

To show that $\alpha_1$ and $\alpha_2$
converge, we claim that it suffices to show unicity of the limit of $\tilde C_Q$.
That is, since for each $\eps$, $\alpha_1$ and $\alpha_2$ are solutions 
of a linear system whose coefficients converge (being functions
of the Green's function derivatives), either in the limit
the linear equation becomes singular (in which case there is non-unicity of the
limit) or it does not, in which case the solution is unique and the 
solutions for finite $\eps$ converge to this solution.

\subsection{Unicity of the limit}
Rather than work on a square region, it suffices to work on the unit disk
with the white hole at the origin: the transformation rules (\ref{transrules1}),
(\ref{transrules2})
allow us to move the result back to the square $Q$.

Let $U$ be the unit disk with 
marked points $b_0,b_1$ on the boundary. 
Let $\tilde U$ be the double cover of $U$ branched over the origin
(the map $f(z)=z^2$ maps $\tilde U$ to $U$).
The lifts to $\tilde U$ of the asymptotic coupling functions $F_0(v,z),F_1(v,z)$
on $U$
have zeros at $z=\pm \sqrt{b_0},\pm \sqrt{b_1}$, poles at $z=\pm v$,
and simple poles at $z=0$. (A priori the poles at the origin 
may have residue $0$,
in which case they are not poles at all). 
These functions also are antisymmetric:
for $v,z\in \tilde U$ we have 
$\tilde F_0(v,z)=-\tilde F_0(v,-z)=\tilde F_0(-v,-z)$ and the same
for $\tilde F_1$.
Furthermore, since $\tilde F_0$ and $\tilde F_1$ are respectively pure imaginary and real
on the boundary of $\tilde U$,
they extend by Schwarz reflection to meromorphic functions on the entire Riemann
sphere, with additional simple poles at $\pm1/\bar v$ and $\infty$
(the reflections of $\pm v$ and $0$).
In particular they are rational functions. Since we know exactly the location
of their (six) poles, and four of their zeros, it remains to find the other
two zeros. The antisymmetry under $z\mapsto -z$ and symmetry under $z\mapsto
1/\bar z$
implies that the two other zeros are on the unit circle: 
otherwise the orbit of the zeros under these two
symmetries would have four elements.
Thus the functions have the form:
$$\tilde F_0(v,z)=c_0\frac{
{\displaystyle\left(
\frac{z}{\sqrt{b_0}}-\frac{\sqrt{b_0}}z\right)
\left(
\frac{z}{\sqrt{b_1}}-\frac{\sqrt{b_1}}z\right)
\left(\frac{z}{b_3}-\frac{b_3}z\right)}}
{(z^2-v^2)(z^{-2}-\bar{v}^2)},$$
$$\tilde F_1(v,z)=ic_1\frac{
{\displaystyle\left(\frac{z}{\sqrt{b_0}}-\frac{\sqrt{b_0}}z\right)\left(
\frac{z}{\sqrt{b_1}}-\frac{\sqrt{b_1}}z\right)\left(\frac{z}{b_4}-\frac{b_4}z\right)}
}
{(z^2-v^2)(z^{-2}-\bar{v}^2)},$$
for real $c_0,c_1$ and 
where $b_3=b_3(v)$, (respectively $b_4=b_4(v)$) must have absolute value $1$.
Now $b_3$ and $b_4$ are 
chosen so that the residues of $\tilde F_0$ and $\tilde F_1$
are real at $z=v$. The constants $c_0=c_0(v),c_1=c_1(v)$ are real
and chosen so that the residue of each function at $z=v$ is $\frac1{\pi}$.

We can explicitly solve for $b_3$ as follows.
The residue of $\tilde F_0$ at $z=v$ is 
$$c_0\frac{\left(\frac{v}{\sqrt{b_0}}-\frac{\sqrt{b_0}}v\right)\left(
\frac{v}{\sqrt{b_1}}-\frac{\sqrt{b_1}}v\right)\left(\frac{v}{b_3}-\frac{b_3}v\right)}
{2v(v^{-2}-\bar{v}^2)}.$$
This is supposed to be real; equating this and its complex conjugate yields
$$\frac{\left(\frac{v}{\sqrt{b_0}}-\frac{\sqrt{b_0}}v\right)\left(
\frac{v}{\sqrt{b_1}}-\frac{\sqrt{b_1}}v\right)\left(\frac{v}{b_3}-\frac{b_3}v\right)}
{2v(v^{-2}-\bar{v}^2)}=
\frac{\left(\bar v\sqrt{b_0}-\frac{1}{\sqrt{b_0}\bar v}\right)\left(
\sqrt{b_1}\bar v-\frac{1}{\bar v\sqrt{b_1}}\right)
\left(b_3\bar v-\frac{1}{\bar vb_3}\right)}
{2\bar v(\bar v^{-2}-v^2)}.$$
Solving for $b_3$ gives
$b_3^2=X_0/\overline{X_0},$ where 
$$X_0=\left(\frac{v}{\sqrt{b_0}}-\frac{\sqrt{b_0}}v\right)\left(
\frac{v}{\sqrt{b_1}}-\frac{\sqrt{b_1}}v\right)v^2+
\left(\bar v\sqrt{b_0}-\frac1{\bar v\sqrt{b_0}}\right)\left(
\bar{v}\sqrt{b_1}-\frac{1}{\bar v\sqrt{b_1}}\right).$$
Similarly one finds $b_4^2=-X_1/\overline{X_1}$, where
$$X_1=-v^2\left(\frac{v}{\sqrt{b_0}}-\frac{\sqrt{b_0}}v\right)\left(
\frac{v}{\sqrt{b_1}}-\frac{\sqrt{b_1}}v\right)+
\left(\bar v\sqrt{b_0}-\frac1{\bar v\sqrt{b_0}}\right)\left(
\bar{v}\sqrt{b_1}-\frac{1}{\bar v\sqrt{b_1}}\right).$$

A lengthy calculation yields
$$\tilde F_+(v,z)=\frac{4z{\displaystyle\left(\frac{z}{\sqrt{b_0}}-\frac{\sqrt{b_0}}z\right)\left( 
\frac{z}{\sqrt{b_1}}-\frac{\sqrt{b_1}}z\right)}}{\pi(z^2-v^2)
{\displaystyle\left(\frac{v}{\sqrt{b_0}}-\frac{\sqrt{b_0}}v\right)
\left(\frac{v}{\sqrt{b_1}}-\frac{\sqrt{b_1}}v\right)}}$$
(which is meromorphic in both $v$ and $z$ as expected)
and
$$\tilde F_-(v,z)=-\frac{4z{\displaystyle\left(\frac{z}{\sqrt{b_0}}-\frac{\sqrt{b_0}}z\right)\left(
\frac{z}{\sqrt{b_1}}-\frac{\sqrt{b_1}}z\right)}}{\pi(1-z^2\bar v^2)
{\displaystyle\left(\bar v\sqrt{b_0}-\frac1{\sqrt{b_0}\bar v}\right)
\left(\bar v\sqrt{b_1}-\frac1{\sqrt{b_1}\bar v}\right)}}$$ 
(anti-meromorphic in $v$ as expected).

The map from $U$ to $\tilde U$ is $f(z)=\sqrt{z}$.
Using the transformation rules (\ref{transrules1}),(\ref{transrules2}) 
we have on the 
original region $U$ 
$$F_+^U(v,z)=\frac{2(z-b_0)(z-b_1)}{\pi(z-v)(v-b_0)(v-b_1)}\sqrt{\frac{v}{z}},$$
$$F_-^U(v,z)=-\frac{2(z-b_0)(z-b_1)}{\pi(1-z\bar v)(\bar vb_0-1)(\bar vb_1-1)}
\sqrt{\frac{\bar v}{z}}.$$
The fact that the transformation rules apply in this case follows
from the fact that
the results $F_\pm^U$ have all the required properties of the coupling
function limits (and are the unique functions with these properties).

\subsection{On the RHP}
The map $f(z)=\frac{z-1}{z+1}$ maps the RHP to the unit disk,
sending $1$ to $0$, $0$ to $-1$ and $\infty$ to $1$.
Let $RHP^\circ$ be the RHP with a `white hole' at $1$.
The limiting coupling functions on $RHP^\circ$
with zeros at $b_5:=f^{-1}(b_0)$ and $b_6:=f^{-1}(b_1)$
are
$$F_+^{RHP^\circ}(v,z)=\frac{2(z-b_5)(z-b_6)}{\pi(z-v)(v-b_5)(v-b_6)}
\sqrt{\frac{v^2-1}{z^2-1}},$$
$$F_-^{RHP^\circ}(v,z)=
\frac{2(z-b_5)(z-b_6)}{\pi(z+\bar v)(\bar v+b_5)(\bar v+b_6)}
\sqrt{\frac{\bar v^2-1}{z^2-1}}.$$
We will assume $b_6=\infty$ ; in this case the formulas on $RHP^\circ$
become
\begin{eqnarray}\label{RHP+2}
F_+^{RHP^\circ}(v,z)=\frac{2(z-b_5)}{\pi(z-v)(v-b_5)}
\sqrt{\frac{v^2-1}{z^2-1}},\\
F_-^{RHP^\circ}(v,z)=-\frac{2(z-b_5)}{\pi(z+\bar v)(\bar v+b_5)}
\sqrt{\frac{\bar v^2-1}{z^2-1}}.\label{RHP-2}
\end{eqnarray}

\subsection{The number of tilings of $Q$}
Let $U$ be the $K\times K$ square $U=[0,K]\times[-K/2,K/2]\subset RHP$.
Suppose $Q=Q_\eps$ approximates $U$ as $\eps\to0$, and suppose
that the holes $b,w,b_0$ of $Q_\eps$, converge to points of the same name
$b,w,b_0\in U$.

Since we are concerned primarily with the growth rate of the LERW,
to simplify the calculations
we will make the further assumption that $b,w\in U$ are within distance
$1$ of the origin, $b_0$ is on the right edge of $U$ and $K$ is large.
Using Lemma \ref{smallj} we can then approximate 
to error $O(1/K)$ the coupling function limits $F_0^U,F_1^U$ 
for points near the origin 
by the (much simpler) coupling function limits on $RHP-\{b,w\}$.

Suppose that $w$ is on the $x$-axis at $x$-coordinate $\alpha>0$ and let $b=\beta i$ 
where $\beta$ is real, $|\beta|<1$.

As in Lemma \ref{2holecomb} 
let $P$ be the region $Q$ with the holes $b,w$ filled in.
Let $S\subset P$ be the chain of horizontal dominos
from the origin to $w$.

Notationally, for a region $A$ let $N(A)$ denote 
the number of tilings of $A$.
The ratio of the number of tilings 
of $Q$ to the number of tilings
of $P$ is a product of three terms
\begin{equation}\label{3prods}
\frac{N(Q)}{N(P)}=\frac{N(Q)}{N(Q-S)}\cdot
\frac{N(Q-S)}{N(P-S)}\cdot\frac{N(P-S)}{N(P)},
\end{equation}
each of which we can now approximate. 

\subsubsection{The first ratio}
The inverse of the first term in (\ref{3prods}), $N(Q-S)/N(Q)$,
is the probability of $S$ occurring in
a tiling of $Q$. This is computed in a similar fashion to
the proof of Theorem \ref{main2}. Let $a_1\ldots,a_m$ be the set of dominos
in the chain $S$. Then the probability of $a_j$ occurring given 
$a_1,\ldots,a_{j-1}$ is given by (see Lemma \ref{nextdom})
$$\frac{Pr(a_j|a_1,\ldots,a_j)}{\sqrt{2}-1}=
1+\left(\frac\pi{\sqrt{2}}\eps F_0^{U_j}(\eps,2\eps)-
1\right)+o(\eps),$$
where $U_j$ is the translate of the region $RHP-[0,2\eps j]$ 
(with a white hole at $w$
and black hole at $b$), translated by $-2\eps j$ so that the tip of the 
cut $(2\eps j)$ is at the origin; $U_j$ corresponds to the polyomino 
$Q_j=Q-\{a_1,\dots,a_{j-1}\}$.
The limiting coupling function $F_0^{U_j}$
on $U_j$ can be computed from (\ref{RHP+2}) and (\ref{RHP-2}).
Indeed, let $t=2\eps j\in[0,\alpha]$; then
the map $$f(z)=\frac{\sqrt{(z+t)^2-t^2}}{\sqrt{\alpha^2-t^2}}$$
maps the region $U_j$ to $RHP$ and sends the singularities $\alpha-t$ to  $1$
and $i\beta-t$ to $b_5=f(i\beta-t)$.

A computation using (\ref{RHP+2}) and (\ref{RHP-2}) gives
\begin{eqnarray*}
\eps F_0^{U_j}(\eps,2\eps)&=&\frac{\eps f'(\eps)}2\left(F_+^{RHP^\circ}(f(\eps),f(2\eps))
+F_-^{RHP^\circ}(f(\eps),f(2\eps))\right)\\
&=&\frac{\sqrt{2}}\pi+\frac{\sqrt{2}(5t^2-\alpha^2)\eps}{
4\pi t(\alpha^2-t^2)}+O(\eps^{3/2}).\end{eqnarray*}

As a consequence 
$$\frac{Pr(a_j|a_1,\ldots,a_j)}{\sqrt{2}-1}=1+\frac{(5t^2-\alpha^2)}{
4t(\alpha^2-t^2)}\eps+o(\eps).$$
When we sum the logs of $Pr(a_j|a_1,\dots,a_{j-1})$ 
for $j$ running from $1$ to $m$ we get 
(see equation (\ref{logprobint}))
$$\log\Pr(S)=m\log(\sqrt{2}-1)+
\int_{2\eps}^{\alpha-2\eps}\frac{(5t^2-\alpha^2)\eps}{4t(\alpha^2-t^2)}\cdot
\frac{dt}{2\eps}+o(\log\eps)+o(1)$$
$$=m\log(\sqrt{2}-1)+\frac18\log\frac1\eps+\frac18\log\alpha+o(1)+
o(\log\eps)$$
where the $o(\log\eps)$ term is independent of $\alpha$ and $\beta$.

Surprisingly, there is no dependence on $\beta$ to first order.
\subsubsection{The second ratio}
Since $b$ and $w$ are on the boundary of the polyomino $P-S$,
the second ratio in (\ref{3prods}) is exactly $|C^{P-S}(w,b)|$, 
where $C^{P-S}$ is the coupling function on $P-S$. 

The coupling function on $P-S$ can be computed using (\ref{transrules1}),
(\ref{transrules2})
and (\ref{RHPF})
and the map $f(z)=\sqrt{(z+\alpha)^2-\alpha^2}$ which maps $RHP-[0,\alpha]$ 
(translated by $-\alpha$
so that the tip of the cut is at the origin) to $RHP$, and $\infty$
to $\infty$. We must compute $\eps F_0^{P-S}(\eps,-\alpha+i\beta)$. We have
$$\eps F_0^{P-S}(\eps,-\alpha+i\beta)=\frac\eps\pi 
f'(\eps)\left(\frac1{f(-\alpha+i\beta)-
f(\eps)}+\frac1{
f(-\alpha+i\beta)+f(\eps)}\right)=\frac{\sqrt{-2\eps\alpha}}{\pi\sqrt{\alpha^2+\beta^2
}}+O(\eps).$$

\subsubsection{The third ratio}
The computation of the third ratio in (\ref{3prods}) is similar to the
first ratio, except that now there are no singularities at $b,w$.
We may use the coupling function on $RHP$, given by (\ref{RHPF}),
to compute the coupling function on $RHP-[0,t]$ (again shifted so that the tip
of the cut is at the origin). The appropriate function is $f(z)=\sqrt{(z+t)^2-t^2}$ 
and we arrive at 
$$\eps F_0^{U_t}(\eps,2\eps)=\frac{\sqrt{2}}\pi-\frac{\sqrt{2}}{4\pi t}\eps+O(\eps^2).$$
The log of the probability of $S$ occurring is then 
$$m\log(\sqrt{2}-1)+\int_{2\eps}^\alpha-\frac\eps{4t}\cdot\frac{dt}{2\eps}$$
$$=m\log(\sqrt{2}-1)-\frac18\log\alpha-\frac18\log\frac1\eps+o(\log\eps)+o(1).$$

\subsubsection{The product}
Now combining these three results, the log of the ratio (\ref{3prods}) is
$$\log(R)= 
-m\log(\sqrt2-1)-\frac18\log\frac1\eps-\frac18\log\alpha+
\log\frac{\sqrt{2\eps\alpha}}{ \pi\sqrt{\alpha^2+\beta^2}} + 
m\log(\sqrt{2}-1)-\frac18\log\alpha-\frac18\log\frac1\eps+o(\log\eps)+o(1)$$
$$=-\frac34\log\frac1\eps+\frac14\log\alpha-\frac12\log(\alpha^2+\beta^2)+
o(\log\eps)+o(1),$$
where the $o(\log\eps)$ is independent of $\alpha,\beta$ (the $o(1)$
depends on $\alpha,\beta$).

This is the log of the ratio of the number of tilings of $Q$
and the number of tilings of $P$.
For finite $K$ there is an additional additive error term $O(1/K)$.

Letting $\alpha+i\beta=re^{i\theta}$ in polar coordinates,
the probability that $w$ is on the LERW from $b$ to
$b_0$ in $P_\eps$ is 
$$\left(\frac{\eps}{r}\right)^{3/4(1+o(1))}\cos(\theta)^{1/4}(1+o(1))$$
where the $o(1)$ in the exponent is independent of $\theta$.

The expected number of edges on
the LERW from $b$ to $b_0$ and which are at distance $\leq R$ from
$b$ is then the integral of 
$\left(\frac{\eps}r\right)^{3/4(1+o(1))}\cos(\theta)^{1/4}$, times the number
of edges per unit area $\frac1{\eps^2}$, times the area form
$r\,dr\,d\theta$, as $r$
runs from $0$ to $R$. This gives $\left(\frac{R}\eps\right)^{5/4(1+o(1))}$. 
This completes the proof of Theorem \ref{lerwthm}.

\section{Open problems}
There are a number of places where our arguments could
stand improving, or where there are interesting avenues for
further research. We list some of the outstanding ones here.
\begin{enumerate}
\item
There remains in Theorem \ref{main2} the error term $ERR(\eps)$
which is somewhat annoying. It seems reasonable to suspect that this
term is in fact $O(1)$ but we don't have any method
of computing this term at present. 
\item 
Is there an extension of Theorem \ref{main2} to the non-simply
connected case, as well as to the case of surfaces of higher genus?
A more general version of Lemma \ref{nextdom} as well as its proof 
should be easy given the coupling function $F_0$.
We don't know at present an easy extension of Lemma \ref{changeinDE} to this case 
though.

\item
Can one find a more natural proof of Lemma \ref{changeinDE}?
There may be a proof using invariance properties of the Schwarzian
derivative but we did not see it.

\item
Theorem \ref{main2} could be generalized to regions with polygonal
boundaries which are horizontal,  vertical, or have slope $\pm1$:
this is because there is an exact formula for the determinant
of the Laplacian on
a triangular region $\{(x,y): x\geq0,~y\geq0,~x+y\leq n\}$, see \cite{KPW}.

More generally, one can ask about the relationship between the term
$-\frac\pi{48}c_2(\eps,U)$ of (\ref{main2form}) and the
$\zeta$-function regularization of the determinant of the Laplacian.
It appears from \cite{OPS} that they are the same up to a multiplicative
constant. Is there a simple explanation for this fact?

\item Is $5/4$ the almost sure growth exponent for the LERW?
The major open question about the LERW is:
Is there a scaling limit of LERW, and if so is it conformally invariant? 
\end{enumerate}

\section{Appendix}
\subsection{Green's function for a slit plane}
We prove here Lemma \ref{Max}.

\begin{proof}
As noted in the first two paragraphs of the proof of Lemma \ref{nextdom},
we must compute the discrete Green's function $G(0,z)$ on $\Z^2-(-\infty,-1]$,
that is, the function of $z$ harmonic on $\Z^2-(-\infty,0]$ with Dirichlet boundary values $0$ 
on the boundary
$(-\infty,-1]$, which satisfies $\Delta G(0,z)= \delta_0(z)$ for $z\in\Z^2-(-\infty,-1]$,
and asymptotically $G(0,z)\to0$ when $|z|\to\infty$.

Let $f_n(z)$ be the harmonic function on $\Z^2-(-\infty,-1]$ whose boundary values
are $0$ for $z\in[-n,-1]$ and $1$ for $z\in(-\infty,-n-1]$.
Let $g_n(z)= f_n(z) - f_{n+1}(z-1).$ By definition, $g_n(z)$ is harmonic
on $\Z^2-(-\infty,0]$ with boundary value $f_{n}(0)$ at $z=0$ and boundary value $0$ for
$z\in(-\infty,-1]$. Therefore $g_n$ is a constant times the desired Green's function
$G(0,z)$. This constant is $1$ over the Laplacian of $g_n$ at $0$, that
is, $-1/(f_{n+1}(0)+f_{n+1}(i)+f_{n+1}(-i))$ 
(note that $f_{n}(z)$ is harmonic at $0$ and $f_{n+1}(z-1)$ is zero at $z=0$ and $z=-1$).
So we have
$$G(0,z) = -\frac{f_n(z)-f_{n+1}(z-1)}{f_{n+1}(0)+f_{n+1}(i)+f_{n+1}(-i)}.$$

The asymptotic values of $f_n(z)$ for large $n$ and $z$ are 
$\frac1\pi\Im\log\frac{1+i\sqrt{z/n}}{1-i\sqrt{z/n}}$.
By Theorem \ref{Kestensthm} applied to the function $f_n(z)$,
we have
$f_n(0)=\lambda_{1,0}\frac{2}{\pi\sqrt{n}}$ and 
$f_n(i)=f_n(-i)=\lambda_{0,1}\frac{\sqrt{2}}{\pi\sqrt{n}}.$
Also when $n\gg|z|\gg1$ we have
$$f_n(z)-f_{n+1}(z-1)=\frac1{\pi\sqrt{nz}}+O(\frac1{z^{3/2}n^{1/2}}).$$
Therefore when $|z|\gg1$, and in the limit as $n\to\infty$, we have
$$G(0,z)=-\left(\frac1{2\lambda_{1,0}+2\sqrt{2}\lambda_{0,1}}\right)\frac1{\sqrt{z}}+O(z^{-3/2}).$$

Scaling everything by $\eps$ (replacing $z$ with $z/\eps$) completes the proof.
\end{proof}

\subsection{Local Riemann mappings}
We prove Lemma \ref{smallj}. 
This is essentially a weaker version of a lemma of Hayman \cite[Lemma 6.6]{Hay}.

First consider the case when $j$ is small and the slit starts
on an edge of $U$. 
For $q>0$ let $g_q$ be the map $g_q(z)=\sqrt{2q z^2+q^2}-q$. Then $g_q$ maps
the RHP injectively onto the translate by $-q$ of $RHP-[0,q]$.
When $q=2\eps j$, $g_q$ is the standard limiting form for the $f_j$; 
we must show that the derivatives of $f_j$ differ from those of $g_{2\eps j}$
by $O(1)$.
There is a constant $C$ (depending on $U$ but independent of $j$)
such that for each $j$ sufficiently small
$g_{2\eps j}^{-1}f_j$ maps a neighborhood of the origin in
RHP to the half-disk $B_C(0)\cap RHP$.
Now $g_{2\eps j}^{-1}f_j$ extends
by Schwarz reflection to an injective map from a ball around the
origin into $B_C(0)$; since the derivative at the origin of $g_{2\eps j}^{-1}f_j$
is $1$ this ball has radius comparable to $C$. 
Since $g_{2\eps j}^{-1}f_j$ is bounded on this
ball, the Schwarz Lemma implies that its derivatives at the origin are all 
bounded as well.
We conclude that $g_{2\eps j}^{-1}f_j=z + z^2O_z(1)$. 

Now $f_j=g_{2\eps j}(z+z^2O_z(1))$ and since $g_{2\eps j}$ is independent
of $U$, the germ of $f_j$ at the origin is independent of $U$ up to $O(1)$.

The same argument with a different $g_q$
works in the case where the cut starts at a corner.
The argument at the end of the cut requires a simple modification.
\hfill{$\square$}

\end{document}